\documentstyle[prd,preprint,aps,amssymb,floats]{revtex} 
\begin{document}


\title{WITH GRAND UNIFICATION SIGNALS IN,\\
	 CAN PROTON DECAY BE FAR BEHIND? \footnote{Invited talk 
   presented at the 
   International School held at Erice Italy (Sept. 2000) and at the 
   Dirac Medalists' symposium held at Trieste, Italy (Nov. 2000). }}
\author{Jogesh C. Pati}
\address{Department of Physics, \\ University of Maryland, \\ 
	College Park  MD 20742, USA.}

\date{\today}

\maketitle
\setlength{\textwidth}{14cm}
\begin{abstract}
It is noted that one is now in possession of a set of
facts, which may be viewed as the {\it matching pieces of a puzzle} ; in
that all
of them can be resolved by just one idea - that is grand unification.
 These include : (i) the observed family-structure, 
(ii) quantization of electric charge, 
(iii) meeting of the three gauge couplings, 
(iv) neutrino oscillations; in particular the mass of $\nu_{\tau}$
(suggested by SuperK), 
(v) the intricate pattern of the masses and mixings of the fermions, 
including the smallness of
$V_{cb}$ and the largeness of $\theta^{osc}_{\nu_{\mu}\nu_{\tau}}$, and 
(vi) the need for $B$-$L$ to implement baryogenesis (via leptogenesis). 
All these pieces fit beautifully together within a
single puzzle board framed by supersymmetric unification, based on 
 SO(10) or a string-unified G(224)-symmetry. 
The one and the most notable piece of the puzzle still missing, 
however, is proton decay.

  A concrete proposal is presented, within a
predictive SO(10)/G(224)-framework, that 
successfully describes the masses and mixings
of all fermions, including the neutrinos - with eight predictions, all in
agreement with observation. Within this
framework, a systematic study of proton decay is carried out, which
pays special attention to its dependence on the fermion masses,
including  the superheavy Majorana masses of the right-handed
neutrinos, and the threshold effects. 
The study (based on prior work and a recent update) 
shows that a conservative upper limit on the
proton lifetime is about (1/2 - 1)$\times10^{34}$ yrs, with
$\overline{\nu}K^{+}$ being the dominant decay mode, and as a
distinctive feature, $\mu^{+}K^{0}$ being prominent. This in turn strongly
suggests that an
improvement in the current sensitivity by a factor of five to
ten (compared to SuperK) ought to reveal proton decay. Otherwise some
promising and remarkably successful ideas on unification would suffer a
major setback.

\end{abstract}
\setlength{\textwidth}{6.5in} 
\section{Introduction}\label{Introduction}
The standard model of particle physics, based on the gauge symmetry 
$SU(2)_L\times U(1)_Y\times SU(3)_C$ \cite{Weinberg,QCD}  
is in excellent agreement 
with observations, at least up to energies of order 100 GeV. Its 
success in turn constitutes a triumph of quantum field theory, 
especially of the notions of gauge invariance, spontaneous 
symmetry breaking, and renormalizability. 
The next step in the unification-ladder is associated with the 
concept of ``grand unification'', which proposes a unity of 
quarks and leptons, and simultaneously of their three basic forces: 
weak, electromagnetic and strong \cite{JCPAS,GG,GQW}. 
This concept was introduced on purely aesthetic grounds, in fact 
{\it before} any of the empirical successes of the standard model was in 
place. It was realized in 1972 that the standard model judged 
on aesthetic merits  
has some major shortcomings \cite{JCPAS,GG}. For example, it puts 
members of a family into five scattered multiplets, 
assigning rather peculiar hypercharge 
quantum numbers to each of them, without however providing a compelling  
reason for doing so. It also does not provide a fundamental reason 
for the quantization of electric charge, and it does not explain 
why the electron and proton possess exactly equal but opposite charges. 
Nor does it explain the co-existence of quarks and leptons, and that of 
the three gauge forces - weak, electromagnetic and strong - with 
their differing strengths. 

The idea of grand unification was postulated precisely to remove these  
shortcomings. It introduces the notion that quarks and 
leptons are members of one family, linked together by a symmetry group 
G, and that the weak, electromagnetic and strong interactions are 
aspects of one force, generated by gauging this symmetry G. The group 
G of course inevitably contains the standard model symmetry 
$G(213)=SU(2)_L\times U(1)_Y\times SU(3)_C$ as a subgroup. Within this 
picture, the observed differences between quarks and leptons and 
those between the three gauge forces are assumed to be low-energy 
phenomena that arise through a spontaneous breaking of the unification 
symmetry G to the standard model symmetry G(213), at a very high energy scale 
$M\gg1 TeV$. As a {\it prediction} of the hypothesis, such differences 
must then disappear and the true unity of quarks and leptons and of the 
three gauge forces should manifest at energies exceeding the scale M.

The second and perhaps the most dramatic prediction of grand unification 
is proton decay. This important process, which would provide the window 
to view physics at truly short distances ($< 10^{-30}$ cm), is  
yet to be seen. Nevertheless, as I will stress in this talk, there has  
appeared over the years an impressive set of facts, favoring the 
hypothesis of grand unification. These include:

{\bf (a)} {\bf The observed family structure :} The five scattered
multiplets of the standard model, belonging to a family, neatly become
parts of a whole ({\it a single multiplet}), with their weak
hypercharges precisely predicted by grand unification. Realization of
this feature calls for an extension of the standard model symmetry
G(213)$\,=\,$SU(2$)_{L}\times$U(1$)_{Y}\times$SU(3$)^{C}$ {\it
minimally} to the symmetry group
G(224)$\,=\,$SU(2$)_{L}\times$SU(2$)_{R}\times$SU(4$)^{C}$ \cite{JCPAS},
which can be extended further into the simple group SO(10)
\cite{SO(10)}, but not SU(5) \cite{GG}. The G(224) symmetry in turn
introduces some additional attractive features (see Sec.\ref{Advantages}), 
including
especially the right-handed (RH) neutrinos ($\nu_{R}$'s) accompanying
the left-handed ones ($\nu_{L}$'s), and $B$-$L$ as a local
symmetry. As we will see, both of these features 
now seem to be needed on empirical grounds.

{\bf (b)} {\bf Meeting of the gauge couplings :} Such a meeting
is found to occur at a scale $M_{X}\approx2\times10^{16}$ GeV, when
the three gauge couplings are extrapolated from their values 
measured at LEP to higher energies, in the context of 
supersymmetry \cite{Langacker}. This dramatic 
phenomenon supports the ideas of both grand unification and
supersymmetry \cite{SUSY}. These in turn may well emerge from a string
theory \cite{String} or M-theory \cite{MTh} 
(see discussion in Sec.\ref{Need}).

{\bf (c)} {\bf Mass of $\bf{\nu_{\tau}\sim1/20}$ eV :} Subject to the
well-motivated assumption of hierarchical neutrino masses, the recent
discovery of atmospheric neutrino-oscillation at SuperKamiokande
\cite{SuperK} suggests  a value for $m(\nu_{\tau})\sim1/20 eV$. It has
been argued (see e.g. Ref. \cite{PatiSuperK}) that
a mass of $\nu_{\tau}$ of this magnitude can be understood very 
simply by utilizing the SU(4)-color relation 
$m(\nu_\tau)_{\mbox{Dirac}}\approx m_{\mbox{top}}$ and the SUSY 
unification scale $M_{X}$, noted above (See Sec.\ref{Mass}).

{\bf (d)} {\bf Some intriguing features of fermion masses and mixings:} 
These include: 
(i) the ``observed'' near equality of the masses of the b-quark and 
    the $\tau$-lepton at the unification-scale 
    (i.e. $m^{0}_{b}\approx m^{0}_{\tau}$) and
(ii) the observed largeness of the $\nu_{\mu}$-$\nu_{\tau}$ oscillation 
     angle 
   ($\sin^{2}2\theta^{\mbox{\scriptsize osc}}_{\nu_{\mu}\nu_{\tau}}\geq0.83$) 
   \cite{SuperK}, together with the smallness of the corresponding quark 
   mixing parameter $V_{cb}(\approx0.04)$ \cite{ParticleDataGroup}.
As shown in recent work by Babu, Wilczek and me \cite{BabuWilczekPati}, 
it turns out that these features and more can be understood remarkably well 
(see discussion in Sec.\ref{Understanding}) within an economical and predictive 
SO(10)-framework based on a minimal Higgs system. The success of this 
framework is in large part due simply to the group-structure of SO(10). 
For most purposes, that of G(224) suffices.

{\bf (e)} {\bf Baryogenesis :} To implement baryogenesis \cite{Sakharov} 
successfully, in the presence of electroweak sphaleron effects
\cite{KuzminRubakov}, which wipe out any baryon excess generated at high
temperatures in the ($B$-$L$)-conserving mode, it has become apparent
that one would need $B$-$L$ as a generator of the underlying symmetry,
whose spontaneous violation at high temperatures would yield, for
example, lepton asymmetry (leptogenesis). The latter in turn is
converted to baryon-excess at lower temperatures by electroweak
sphalerons. This mechanism, it turns out, yields even quantitatively
the right magnitude for baryon excess \cite{LeptoB}. The need for
$B$-$L$, which  is a generator of SU(4)-color, again points to the
need for G(224) or SO(10) as an effective symmetry near the
unification-scale $M_{X}$.

The success of each of these five features (a)-(e) seems to be
non-trivial. Together they make a strong case for both
supersymmetric grand unification and simultaneously for the
G(224)/SO(10)-route to such unification, as being relevant to
nature at short distances. However, despite these successes, as 
long as proton decay
remains undiscovered, the hallmark of grand unification - that is {\it
quark-lepton transformability} - would remain unrevealed.

The relevant questions in this regard then are : What is the predicted
range for the lifetime of the proton - in particular an upper limit -
within the empirically favored route to unification mentioned above?
What are the expected dominant decay modes within this route? Are
these predictions compatible with current lower limits on proton
lifetime mentioned above, and if so, can they still be tested at the
existing or possible near-future detectors for proton decay?

Fortunately, we are in a much better position to answer these
questions now, compared to a few years ago, because meanwhile we have
learnt more about the nature of grand unification. As noted above (see
also Sec.\ref{Advantages} and Sec.\ref{Mass}), 
the neutrino masses and the meeting of the gauge
couplings together seem to select out the supersymmetric
G(224)/SO(10)-route to higher unification. The main purpose of my talk
here will therefore be to address the questions raised above, in the
context of this route. For the sake of comparison, however, I will
state the corresponding results for the case of supersymmetric SU(5)
as well.

My discussion will be based on a recent study of proton decay by Babu,
Wilczek and me \cite{BabuWilczekPati} and an update of the same 
as presented here. Relative to other analysis, this study has 
three distinctive features: 

{\bf (a)} It systematically takes into account the link that exists
between proton decay and the masses and mixings of all fermions,
including the neutrinos. 

{\bf (b)} In particular, in addition to the contributions from the
so-called ``standard'' $d=5$ operators \cite{Sakai} 
(see Sec.\ref{Expectations}), 
it includes those from a {\it new} set of $d=5$ operators, related 
to the Majorana masses of the RH neutrinos \cite{BPW1}. These latter are
found to be as important as the standard ones. 

{\bf (c)} The work also incorporates  GUT-scale threshold effects,
 which arise because of mass-splittings between the components of the 
 SO(10)-multiplets, and lead to differences between the three gauge
 couplings.

Each of these features turn out to be {\it crucial} to gaining a 
reliable insight into the nature of proton decay. 
Our study shows that the inverse decay rate
for the $\overline{\nu}K^{+}$-mode, which is dominant, is less than
about $5\times10^{33}$ yrs for the case of MSSM embedded in SO(10). 
This upper bound is obtained by making
generous allowance for uncertainties in the matrix element and the
SUSY-spectrum. Typically, the lifetime should of course be less than
this bound. 

Proton decay is studied also for the case of the extended 
supersymmetric standard model (ESSM), that has been proposed a few 
years ago \cite{BabuJi} on theoretical grounds, pertaining to the 
issues of string-unification and dilaton stabilization 
(see Sec.\ref{Expectations} and the appendix). This case adds an extra 
pair of vector-like families at the TeV-scale, transforming as 
$16+\overline{16}$ of SO(10), to the MSSM spectrum. While the case 
of ESSM is fully compatible with both neutrino-counting at LEP 
and precision electroweak tests, it can of course be tested directly 
at the LHC. Our study shows that, with the inclusion of only the 
standard d=5 operators (defined in Sec.\ref{Expectations}), ESSM, 
embedded in SO(10), can quite plausibly lead to proton lifetimes 
in the range of $10^{33} - 10^{34}$ years, for nearly central 
values of the parameters pertaining to the SUSY-spectrum and 
the matrix element. Allowing for a wide variation of the parameters, 
owing to the contributions from both the standard and the neutrino 
mass-related d=5 operators (discussed in Sec.\ref{Expectations}), proton 
lifetime still gets bounded above by about $10^{34}$ years, even for 
the case of ESSM, embedded in SO(10) or a string - G(224).

For either MSSM and ESSM, due to contributions from the new
operators, the $\mu^{+}K^{0}$-mode is found to be prominent, with a
branching ratio typically in the range of 10-50\%. By contrast,
minimal SUSY SU(5), for which the new operators are absent, would lead
to branching ratios $\leq10^{-3}$ for this mode. 

Thus our study of
proton decay, correlated with fermion masses, strongly suggests that
discovery of proton decay should be imminent. In fact,one expects 
that at least candidate events  should be observed in the near future
already at SuperK. However, allowing for the possibility that the
proton lifetime may well be closer to the upper bound stated above, a
next-generation detector providing a net gain in sensitivity in proton
decay-searches by a factor of 5-10, compared to SuperK, would
certainly be needed not just to produce proton-decay events, but also
to clearly distinguish them from the background. It would of course
also be essential  to study the branching ratios of certain
sub-dominant but crucial decay modes, such as the $\mu^{+}K^{0}$. The
importance of such improved sensitivity, in the light of the successes
of supersymmetric grand unification, is emphasized at the end.

\section{Advantages of the Symmetry G(224) as a Step to Higher 
	 Unification}\label{Advantages}
\noindent As mentioned in the introduction, the hypothesis of grand 
unification was introduced to remove some of the conceptual shortcomings 
of the standard model (SM). To illustrate the advantages of an early 
suggestion in this regard, consider the five standard model multiplets 
belonging to the electron-family as shown : 
\begin{eqnarray}
\left(\begin{array}{ccc}{u_{r}}\,\,\,\,{u_{y}}\,\,\,\,{u_{b}}\\{d_{r}}\,\,\,\,{d_{y}}\,\,\,\,{d_{b}}\end{array}\right)^{\frac{1}{3}}_{L}\,;\,\,
\left(\begin{array}{ccc}{u_{r}}\,\,\,\,{u_{y}}\,\,\,\,{u_{b}}\end{array}\right)^{\frac{4}{3}}_{R}\,;\,\,
\left(\begin{array}{ccc}{d_{r}}\,\,\,\,{d_{y}}\,\,\,\,{d_{b}}\end{array}\right)^{-\,\frac{2}{3}}_{R}\,;\,\,
\left(\begin{array}{c}{\nu_{e}}\\{e^{-}}\end{array}\right)^{-\,1}_{L}\,;\,\,
\left(e^{-}\right)^{-\,2}_{R}\,.
\label{e1}
\end{eqnarray}
Here the superscripts denote the respective weak hypercharges $Y_{W}$
(where $Q_{em}=I_{3L}+Y_{W}/2$) and the subscripts L and R denote the
chiralities of the respective fields. If one asks : how one can put
these five multiplets into just one multiplet, the answer turns out to
be simple and unique. As mentioned in the introduction, the minimal
extension of the SM symmetry G(213) needed, to achieve this goal, is
given by the gauge symmetry \cite{JCPAS} :
\begin{eqnarray}
\mbox{G(224)}\,=\,\mbox{SU(2})_{L}\times \mbox{SU(2})_{R}\times \mbox{SU(4})^{C}\,.
\label{e2}
\end{eqnarray}
Subject to left-right discrete symmetry ($L\leftrightarrow R$), which
is natural to G(224), all members of the electron family fall into the
neat pattern :
\begin{eqnarray}
F^{e}_{L,\,R}\,=\,\left[\begin{array}{cccc}{u_{r}}\,\,\,\,{u_{y}}\,\,\,\,{u_{b}}\,\,\,\,{\nu_{e}}\\{d_{r}}\,\,\,\,{d_{y}}\,\,\,\,{d_{b}}\,\,\,\,{{e}^{-}}\end{array}\right]_{L,\,R}
\label{e3}
\end{eqnarray}
The multiplets $F^{e}_{L}$ and $F^{e}_{R}$ are left-right conjugates
of each other and transform respectively as (2,1,4) and (1,2,4) of
G(224); likewise for the muon and the tau families. Note that the
symmetries SU(2$)_{L}$ and SU(2$)_{R}$ are just like the familiar
isospin symmetry, except that they operate on quarks and well as
leptons, and distinguish between left and right chiralities. The left
weak-isospin SU(2$)_{L}$ treats each column of $F^{e}_{L}$ as a
doublet; likewise SU(2$)_{R}$ for $F^{e}_{R}$. The symmetry
SU(4)-color treats each row of $F^{e}_{L}$ {\it and} $F^{e}_{R}$ as a
quartet; {\it thus lepton number is treated as the fourth color}. Note also
that postulating either SU(4)-color or SU(2$)_{R}$ forces one to
introduce a right-handed neutrino ($\nu_{R}$) for each family as a
singlet of the SM symmetry. {\it This requires that there be sixteen
two-component fermions in each family, as opposed to fifteen for the
SM}. The symmetry G(224) introduces an elegant charge formula : 
\begin{eqnarray}
Q_{em}\,=\,I_{3L}\,+\,I_{3R}\,+\,\frac{B\,-\,L}{2}
\label{e4}
\end{eqnarray}
expressed in terms of familiar quantum numbers $I_{3L}$, $I_{3R}$ and
$B$-$L$, which applies to all forms of matter (including quarks and
leptons of all six flavors, gauge and Higgs bosons). Note that the
weak hypercharge given by $Y_{W}/2=I_{3R}\,+\,\frac{B\,-\,L}{2}$ is
now completely determined for all members of the family. The values of
$Y_{W}$ thus obtained precisely match the assignments shown in
Eq. (\ref{e1}). Quite clearly, the charges $I_{3L}$, $I_{3R}$ and
$B$-$L$, being generators respectively of SU(2$)_{L}$, SU(2$)_{R}$ and
SU(4$)^{c}$, are quantized; so also then is the electric charge
$Q_{em}$. 

In brief, the symmetry G(224) brings some attractive features to
particle physics. These include :\\  (i) Unification of all 16
members of a family within one left-right self-conjugate multiplet;\\
(ii) Quantization of electric charge, with a reason for the fact that  
$Q_{\mbox{\scriptsize{electron}}}=-Q_{\mbox{\scriptsize{proton}}}$
\\  (iii) Quark-lepton
unification (through SU(4) color);\\  (iv) Conservation of parity at a
fundamental level \cite{JCPAS,RNMJCP};\\  (v) Right-handed neutrinos
($\nu_{R}'s$) as a compelling feature; and\\  (vi) $B$-$L$ as a local
symmetry.\\  As mentioned in the introduction, the two distinguishing
features of G(224) - i.e. the existence of the RH neutrinos and
$B$-$L$ as a local symmetry - now seem to be needed on empirical
grounds. Furthermore, SU(4)-color provides simple relations between 
the masses of quarks and leptons, especially of those in the third 
family. As we will see in Secs.\ref{Mass} and \ref{Understanding}, 
these are in good accord with observations. 

Believing in a complete unification, one is led to view the G(224)
symmetry as part of a bigger symmetry, which itself may have its
origin in an underlying theory, such as string theory. In this
context, one may ask : Could the effective symmetry below the string
scale in four dimensions (see Sec.\ref{Need}) be as small as just the SM
symmetry G(213), even though the latter may have its origin in a bigger
symmetry, which lives  only in higher
dimensions? I will argue in Sec.\ref{Mass} that the data on neutrino masses
and the need for baryogenesis provide an answer to the contrary,
suggesting that it is the effective symmetry in four
dimensions, below the string scale, which must {\it minimally} contain
either G(224) or a close relative
G(214)$\,=\,$SU(2$)_{L}\times$I$_{3R}\times$SU(4$)^{C}$. 

One may also ask : does the effective four dimensional symmetry  have
to be any bigger than G(224) near the string scale? In preparation for
an answer to this question, let us recall that the smallest simple
group that contains the SM symmetry G(213) is SU(5) \cite{GG}. It has
the virtue of demonstrating how the main ideas of grand unification,
including unification of the gauge couplings, can be
realized. However, SU(5) does not contain G(224) as a subgroup. As
such, it does not possess some of the advantages listed above. In
particular, it does not contain the RH neutrinos as a compelling
feature, and $B$-$L$ as a local symmetry. Furthermore, it splits
members of a family into two multiplets : $\overline{5}+10$. 

By contrast, the symmetry SO(10) has the merit, relative to SU(5),
that it contains G(224) as a subgroup, and thereby retains all the
advantages of G(224) listed above. (As a historical note, it is worth
mentioning that these advantages had been motivated on aesthetic 
grounds  through the symmetry G(224) \cite{JCPAS}, and {\it all} 
the ideas of higher unification were in place \cite{JCPAS,GG,GQW}, 
before it was noted that G(224)(isomorphic to SO(4)$\times$SO(6)) 
embeds nicely into SO(10) \cite{SO(10)}). Now, {\it SO(10) even preserves 
the 16-plet  family-structure of G(224) without a
need for any extension}. By contrast, if one extends G(224) to the
still higher symmetry E$_{6}$ \cite{E6}, the advantages (i)-(vi) are
retained, but in this case, one must extend the family-structure from
a 16 to a 27-plet, by postulating additional fermions. In this sense,
there seems to be some advantage in having the effective symmetry
below the string scale to be minimally G(224) (or G(214)) and
maximally no more than SO(10). I will compare the relative advantage
of having either a string-derived G(224) or a string-SO(10), in the
next section. First, I discuss the implications of the data on
coupling unification. 

\section{The Need for Supersymmetry : MSSM versus String Unifications}
\label{Need}
It has been known for some time that the precision measurements of the
standard model coupling constants (in particular $\sin^{2}\theta_{W}$)
at LEP put severe constraints on the idea of grand unification. Owing
to these constraints, the non-supersymmetric minimal SU(5), and for
similar reasons, the one-step breaking minimal non-supersymmetric
SO(10)-model as well, are now excluded \cite{LangPolonski}. But the situation
changes radically if one assumes that the standard model is replaced
by the minimal supersymmetric standard model (MSSM), above a threshold
of about 1 TeV. In this case, the three gauge couplings are found to
meet \cite{Langacker}, to a very good approximation, barring a few 
percent discrepancy which can be attributed to threshold corrections 
(see Appendix). Their scale of meeting is given by
\begin{eqnarray}
M_{X}\approx 2\times10^{16}\,\mbox{GeV\,\,\,\,(MSSM or SUSY\,\,SU(5))}
\label{e5}
\end{eqnarray}

This dramatic meeting of the three gauge couplings, or equivalently
the agreement of the MSSM-based prediction of
$\sin^{2}\theta_{W}(m_{Z})_{\mbox{Th}}=0.2315\pm0.003$ \cite{no26}
with the observed value of
$\sin^{2}\theta_{W}(m_{Z})=0.23124\pm0.00017$ \cite{ParticleDataGroup}, 
provides a strong support for the ideas of both grand unification and
supersymmetry, as being relevant to physics at short distances. 

In addition to being needed for achieving coupling unification there 
is of course an independent motivation for low-energy supersymmetry 
- i.e. for the existence of SUSY partners of the standard model 
particles with masses of order 1 TeV. This is because it protects the 
Higgs boson mass from getting large quantum corrections, which would 
(otherwise)  
arise from grand unification and Planck scale physics. It thereby 
provides at least a technical resolution of the so-called 
gauge-hierarchy problem. In 
this sense low-energy supersymmetry seems to be needed for the 
consistency of the hypothesis of grand unification. Supersymmetry is 
of course also needed for the consistency of string theory. And most 
important, low-energy supersymmetry can be tested at the LHC, and 
possibly at the Tevatron. 

The most straightforward interpretation of the observed meeting of the
three gauge couplings and of the scale $M_{X}$, is that a supersymmetric
grand unification symmetry (often called GUT symmetry), like SU(5) or
SO(10), breaks spontaneously at $M_{X}$ into the standard model
symmetry G(213).

Even if supersymmetric grand unification may well be a good effective 
theory below a certain scale $M \gtrsim M_X$, it ought to have its 
origin within an underlying theory like string/M theory. Such a theory 
is needed to unify all the forces of nature including gravity, and to 
provide a good 
quantum theory of gravity. It is also needed to provide a rationale 
for the existence of flavor symmetries (not available within grand 
unification), which distinguish between the three families 
and can resolve certain 
naturalness problems including those associated with inter-family 
mass hierarchy. 

In the context of string or M theory, an alternative interpretation 
of the observed meeting of the gauge couplings is however 
possible. This is because, even if the effective symmetry in
four dimensions emerging from a higher dimensional string theory is
non-simple, like G(224) or G(213), string theory can still ensure
familiar unification of the gauge couplings at the string scale. In
this case, however, one needs to account for the small mismatch
between the MSSM unification scale $M_{X}$ (given above), and the
 string unification scale, given by $M_{st}\approx
g_{st}\times5.2\times10^{17}$ GeV $\approx 3.6\times10^{17}$ GeV 
(Here we have put $\alpha_{st}=\alpha_{GUT}(\mbox{MSSM})\approx0.04$)
\cite{GinsporKap}. Possible resolutions of this mismatch have been
proposed. These include : (i) utilizing the idea of {\it
string-duality} \cite{WittenDual} which  allows a lowering of $M_{st}$
compared to the value shown above, or alternatively   (ii) the idea of
a {\it semi-perturbative} unification that assumes the existence of
two vector-like families, transforming as $(16+\overline{16})$ of SO(10), 
with masses of order one TeV \cite{BabuJi}. The latter
raises $\alpha_{GUT}$ to about 0.25-0.3 and simultaneously $M_{X}$, in
two loop, to about $(1/2-2)\times10^{17}$ GeV. (Other
mechanisms resolving the mismatch are reviewed in 
Ref.\cite{DienesJCP}). In practice, a combination of the two mechanisms
mentioned above may well be relevant.
\footnote{
  I have in mind the
  possibility of string-duality \cite{WittenDual} lowering $M_{st}$ 
  for the case of semi-perturbative unification (for which
  $\alpha_{st}\approx$0.25, and thus, without the use of string-duality,
  $M_{st}$ would be about $10^{18}$ GeV) to a value of about
  (1-2)$\times10^{17}$ GeV (say), and semi-perturbative unification
  \cite{BabuJi}
  raising the MSSM value of $M_{X}$ to about 5$\times10^{16}$ GeV$
  \approx$ $M_{st}$(1/2 to 1/4) (say). In this case,
  an intermediate symmetry like G(224) emerging at $M_{st}$ would be
  effective only within the short gap between $M_{st}$ and $M_{X}$,
  where it would break into G(213). Despite this short gap, one would
  still have the benefits of SU(4)-color that are needed to
  understand neutrino masses (see sec.4). At the same time, since the
  gap is so small, the couplings of G(224), unified at $M_{st}$ would
  remain essentially so at $M_{X}$, so as to match with the ``observed''
  coupling unification, of the type suggested in Ref.\cite{BabuJi}.
}

While the mismatch can thus quite plausibly be removed for a non-GUT
string-derived symmetry like G(224) or G(213), a GUT symmetry like
SU(5) or SO(10) would have an advantage in this regard because it
would keep the gauge couplings together between $M_{st}$ and $M_{X}$
(even if $M_{X}\sim M_{st}/20$), and thus  not even encounter the
problem of a mismatch between the two scales. A supersymmetric
GUT-solution (like SU(5) or SO(10)), however, has a possible
disadvantage as well, because it needs certain color triplets to
become superheavy by the so-called doublet-triplet splitting mechanism
(see Sec.\ref{Expectations} and Appendix), in order to avoid the problem of rapid
proton decay. However, no such mechanism has emerged yet, in string
theory, for the GUT-like solutions \cite{StringGUT}.

Non-GUT string solutions, based on symmetries like G(224) or G(2113)
for example, have a distinct advantage in this regard, in that the
dangerous color triplets, which would induce rapid proton decay, are
often naturally projected out for such solutions
\cite{Antoniadis,FaraggiHalyo}. Furthermore, the non-GUT solutions invariably
possess new ``flavor'' gauge symmetries, which distinguish between
families. These symmetries are immensely helpful in explaining
qualitatively the observed fermion mass-hierarchy (see
e.g. Ref. \cite{FaraggiHalyo}) and resolving the so-called naturalness
problems of supersymmetry such as those pertaining to the issues of
squark-degeneracy \cite{FaraggiJCP}, CP violation 
\cite{BabuJCP} and quantum gravity-induced rapid proton decay 
\cite{JCPProton}.

Weighing the advantages and possible disadvantages of both, it seems
hard at present to make a priori a clear choice between a GUT versus
a non-GUT string-solution. As expressed elsewhere \cite{JCPRef}, it
therefore seems prudent to keep both options open and pursue their
phenomenological consequences. Given the advantages of G(224) or
SO(10) in the light of the neutrino masses (see Secs.\ref{Advantages} 
and \ref{Mass}), I will
thus proceed by assuming that either a suitable G(224)-solution with a
mechanism of the sort mentioned above, or a realistic SO(10)-solution
with the needed doublet-triplet mechanism, will emerge from string
theory. We will see that with this broad assumption, an economical and
predictive framework emerges, which  successfully accounts for a host of
observed phenomena, and makes some crucial testable predictions. 
Fortunately, it will turn out that there are many similarities between the
predictions of a string-unified G(224) and SO(10) frameworks, 
not only for the neutrino and the charged fermion masses, but also for
proton decay. I next discuss the implications of the mass of
$\nu_{\tau}$ suggested by the SuperK data.

\section{Mass of $\nu_\tau$: Evidence In Favor of the G(224) Route}
\label{Mass}
One can obtain an estimate for the mass of $\nu_{L}^{\tau}$ in the
context of G(224) or SO(10) by using the following three steps
(see e.g.Ref.\cite{PatiSuperK}): 
 
(i) Assume that B$-$L and $I_{3R}$, contained in a string-derived
G(224) or SO(10), break near the unification-scale:  
\begin{eqnarray} 
M_X\sim2\times10^{16}\,\mbox {GeV}\,, 
\label{e6}
\end{eqnarray} 
through VEVs of Higgs multiplets of the type suggested by
string-solutions - i.e. $\langle(1,2,4)_H\rangle$ for
G(224) or $\langle\overline{16}_{H}\rangle$ for  SO(10), as opposed to
$126_H$ which seems to be unobtainable (at least) in weakly interacting 
string theory \cite{DienesRussell}.  
In the process, the RH neutrinos  ($\nu_{R}^{i}$), which are
singlets of the standard model, can and  generically will acquire
superheavy Majorana masses of the type
$M_{R}^{ij}\,\nu_R^{iT}\,C^{-1}\,\nu_R^{j}$, by utilizing the VEV of
$\langle{\overline{16}}_{H}\rangle$ and effective  couplings of the
form: 
\begin{eqnarray} 
{\cal L}_M\,(SO(10))\,=\,f_{ij}\,\,16_{i}\cdot16_{j}\,\,\overline{16}_{H}\,\cdot\overline{16}_{H}/M
+ h.c. 
\label{e7}
\end{eqnarray} 
 
A similar expression holds for G(224). Here $i,j=1,2,3$, correspond
respectively to $e,\,\mu$ and $\tau$ families.  Such gauge-invariant
non-renormalizable couplings might be expected to be induced by
Planck-scale  physics, involving quantum gravity or stringy effects
and/or tree-level exchange of superheavy  states, such as those in the
string tower.  With $f_{ij}$ (at least   the largest among them) being
of order unity, we would thus expect M to lie   between
$M_{Planck}\approx2\times10^{18}$ GeV and
$M_{string}\approx4\times10^{17}$ GeV. Ignoring for the present
off-diagonal mixings (for simplicity), one thus obtains 
\footnote{
  The effects of neutrino-mixing and of possible choice of 
  $M=M_{string}\approx4\times 10^{17}$ GeV (instead of $M=M_{Planck}$) 
  on $M_{3R}$ are considered in Ref. \cite{BabuWilczekPati}.
}:
\begin{eqnarray} 
M_{3R}\,\approx\,\frac{f_{33}\langle\overline{16}_{H}\rangle^{2}}{M} 
\,\approx\,f_{33}\,(2\times10^{14}\,\mbox{GeV})\,\rho^{2}\,(M_{Planck}/M) 
\label{e8}
\end{eqnarray} 
 
This is the Majorana mass of the RH tau neturino. Guided by the 
value of $M_{X}$, we have substituted 
$\langle\overline{16}_{H}\rangle=(2\times10^{16}\,\mbox{GeV})\,\rho$ 
,with $\rho\approx1/2$ to 2(say).  
 
(ii) Now using SU(4)-color and the
Higgs  multiplet $(2,2,1)_{H}$ of G(224) or equivalently $10_{H}$ of
SO(10), one  obtains the relation $m_{\tau}(M_{X}) = m_{b}(M_{X})$,
which is known to be  successful. Thus, there is a good reason to
believe that the third family gets its masses primarily from the
$10_{H}$ or equivalently $(2,2,1)_{H}$ (see sec.5). In turn, this implies: 
\begin{eqnarray} 
m(\nu^{\tau}_{Dirac})\,\approx\,m_{top}(M_{X})\,\approx\,(100\,\mbox{-}\,120)\,\mbox{GeV} 
\label{e9}
\end{eqnarray} 
Note that this relationship between the Dirac mass of  the tau-neutrino 
and the top-mass is special to SU(4)-color.  It does not emerge in 
SU(5). 
 
(iii) Given the superheavy Majorana masses of the RH neutrinos as well 
as the Dirac masses as above, the see-saw mechanism \cite{SeeSaw} 
yields naturally light masses for the LH neutrinos.  For $\nu_{L}^{\tau}$ 
(ignoring flavor-mixing), one thus obtains, using Eqs.(\ref{e8}) and (\ref{e9}),  
\begin{eqnarray} 
m(\nu^{\tau}_{L})\,\approx\,\frac{m(\nu^{\tau}_{Dirac})^2}{M_{3R}}\,\approx\,[(1/20)\,\mbox{eV}\,(1\,\mbox{-}\,1.44)/f_{33}\,\rho^{2}]\,(M/M_{Planck}) 
\label{e10}
\end{eqnarray} 
 
Now, assuming the hierarchical pattern $m(\nu^{e}_{L})\ll
m(\nu^{\mu}_{L})\ll m(\nu^{\tau}_{L})$, which is suggested by the see-saw
mechanism, and further that the SuperK observation represents
$\nu_{L}^{\mu}-\nu_{L}^{\tau}$ (rather than $\nu_{L}^{\mu}-\nu_{X}$)
oscillation, the observed  $\delta
m^2\approx1/2(10^{-2}\,\mbox{-}\,10^{-3})\,$eV$^{2}$ corresponds  to
$m(\nu_{L}^{\tau})\approx$ (1/15 - 1/40) eV. It seems  {\it
truly remarkable} that the expected magnitude of $m(\nu_{L}^{\tau})$,
given by Eq.(\ref{e10}), is just about what is suggested by the SuperK
data, if $f_{33}\,\rho^2(M_{Planck}/M)\approx$ 1.3 to 1/2.  Such a 
range for $f_{33}\,\rho^2(M_{Planck}/M)$ seems most plausible and
natural (see discussion in Ref. \cite{PatiSuperK}). Note that
the estimate (\ref{e10}) crucially depends upon the supersymmetric
unification scale, which provides a value for $M_{3R}$,
as well as on SU(4)-color that yields $m(\nu^{\tau}_{Dirac})$.{\it The 
agreement between the expected and the SuperK results thus clearly 
favors supersymmetric unification, and in the string theory context, it 
suggests that the effective symmetry below the string-scale should
contain SU(4)-color}. Thus, minimally this effective symmetry should be 
either G(214) or G(224), and maximally as big as SO(10), if not E$_{6}$. 
 
By contrast, if SU(5) is regarded as either a fundamental symmetry or as
the effective symmetry below the string scale, there would be no
compelling reason based on symmetry alone, to introduce a $\nu_{R}$,
because it is a singlet of SU(5).  Second, even if one did introduce
$\nu^{i}_{R}$ by hand, their Dirac masses, arising from the coupling
$h^{i}\,\overline{5}_{i}\langle 5_H\rangle\nu^{i}_{R}$, would be
unrelated to the up-flavor  masses and thus rather arbitrary (contrast
with Eq. (\ref{e9})).  So also  would be the Majorana masses of the
$\nu^{i}_{R}$'s, which are  SU(5)-invariant, and thus can be even of
order string scale . This would give
$m(\nu^{\tau}_{L})$ in gross conflict  with the observed value. 

Before passing to the next section, it is worth noting that the mass of
$\nu_{\tau}$ suggested by SuperK, as well as the observed value of
$\sin^{2}\theta_{W}$ (see Sec.\ref{Need}), provide 
valuable insight into the nature
of GUT symmetry breaking. They both favor the case of a {\it single-step
breaking} (SSB) of SO(10) or a string-unified G(224) symmetry at a scale
of order $M_{X}$, into the standard model symmetry G(213), as opposed to
that of a multi-step breaking (MSB). The latter would correspond, for
example, to SO(10) (or G(224)) breaking at a scale $M_{1}$ into G(2213),
 which in turn breaks at a scale $M_{2}<< M_{1}$ into G(213).
 One reason why the case of single-step  breaking is favored over that of
multi-step breaking is that the latter can accommodate but not really
predict $\sin^{2}\theta_{W}$, whereas the former predicts the same
successfully. Furthermore, since the Majorana mass of $\nu^{\tau}_{R}$
arises arises only after $B-L$ and $I_{3R}$ break, it would be given, for
the case of MSB, by $M_{3R}\sim f_{33}(M_{2}^{2}/M)$, where $M\sim M_{st}$
(say). If $M_{2}\ll M_{X}\sim 2\times10^{16}$ GeV, and M $>M_{X}$, one
would obtain too low a value ($<< 10^{14}$ GeV) for $M_{3R}$ (compare with
Eq.(8)), and thereby too large a value for $m(\nu^{\tau}_{L})$, compared to
that suggested by SuperK. By contrast, the case of SSB yields the right
magnitude for $m(\nu_{\tau})$ (see Eq. (10)).

Thus the success of the
result on $m(\nu_{\tau})$ discussed above not only favors the symmetry
G(224) or SO(10), but also clearly suggests that $B-L$ and $I_{3R}$ break
near the conventional GUT scale $M_{X}\sim 2\times 10^{16}$ GeV, rather
than at an intermediate scale $<< M_{X}$. In other words, the observed
values of both $\sin^{2}\theta_{W}$ and $m(\nu_{\tau})$ favor
only {\it the simplest pattern of symmetry-breaking}, for which SO(10) or a
string-derived G(224) symmetry breaks in one step to the
standard model symmetry, rather than in multiple steps. It is of course
only this simple
pattern of symmetry breaking that would be rather restrictive as regards
its predictions for proton decay (to be discussed in Sec.\ref{Expectations}). 
I next discuss the problem of understanding the masses and mixings of all fermions.       

\section{Understanding Fermion Masses and Neutrino Oscillations in SO(10)}
\label{Understanding}

Understanding the masses and mixings of all quarks and charged
leptons, in conjunction with those of the neutrinos, is a goal worth
achieving by itself. It also turns out to be essential for the study
of proton decay. I therefore present first a recent attempt in this
direction, which seems most promising \cite{BabuWilczekPati}. A few 
guidelines would prove to be helpful in this regard. The first of these 
is motivated by the desire for economy and the rest by data. 

{\bf 1) Hierarchy Through Off-diagonal Mixings :} Recall earlier
attempts \cite{Weinbergetal} that attribute hierarchical masses of the first
two families to mass matrices of the form :
\begin{eqnarray}
M\,=\,\left(\begin{array}{cc}{0}\,\,\,\,\,\,\,\,\,\,\,\,{\epsilon}\\{\epsilon}\,\,\,\,\,\,\,\,\,\,\,\,{1}\end{array}\right)\,m^{(0)}_{s}\,,
\label{e11}
\end{eqnarray}
for the $(d,s)$ quarks, and likewise for the $(u,c)$ quarks. Here
$\epsilon\sim1/10$. The hierarchical patterns in Eq. (\ref{e11}) can
be ensured by imposing a suitable flavor symmetry which distinguishes
between the two families (that in turn may have its origin in string
theory (see e.g. Ref \cite{FaraggiHalyo}). Such a pattern has the virtues that
(a) it yields a hierarchy that is much larger than the input
parameter $\epsilon$ : $(m_{d}/m_{s})\approx\epsilon^{2}\ll\epsilon$,
and (b) it leads to an expression for the cabibbo
angle :
\begin{eqnarray}
\theta_{c}\approx\bigg|\sqrt{\frac{m_{d}}{m_{s}}}\,-\,e^{i\phi}\,\sqrt{\frac{m_{u}}{m_{c}}}\,\bigg|\,,
\label{e12}
\end{eqnarray}
which is rather successful. Using $\sqrt{m_{d}/m_{s}}\approx 0.22$ and
$\sqrt{m_{u}/m_{c}}\approx
0.06$, we see that Eq. (\ref{e12}) works to within about $25\%$ for
any value of the phase $\phi$. Note that the square root formula (like
$\sqrt{m_{d}/m_{s}}$) for the relevant mixing angle arises because of
the symmetric form of $M$ in Eq. (\ref{e11}), which in turn is ensured
if the contributing Higgs is a 10 of SO(10). A generalization of the
pattern in Eq. (\ref{e11}) would suggest that the first two families
(i.e. the $e$ and the $\mu$) receive masses primarily through
their mixing with the third family $(\tau)$, with $(1,3)$ and $(1,2)$
elements being smaller than the $(2,3)$; while  $(2,3)$ is smaller than
the
$(3,3)$. We will follow this guideline, except for the modification
noted below.

{\bf 2) The Need for an Antisymmetric Component :} Although the
symmetric hierarchical matrix in Eq. (\ref{e11}) works well for the
first two families, a matrix of the same form fails altogether to
reproduce $V_{cb}$, for which it yields :
\begin{eqnarray}
V_{cb}\approx\bigg|\sqrt{\frac{m_{s}}{m_{b}}}\,-\,e^{i\chi}\,\sqrt{\frac{m_{c}}{m_{t}}}\,\bigg|\,.
\label{e13}
\end{eqnarray}
Given that $\sqrt{m_{s}/m_{b}}\approx 0.17$ and
$\sqrt{m_{c}/m_{t}}\approx 0.0.06$, we see that Eq. (\ref{e13}) would
yield $V_{cb}$ varying between 0.11 and 0.23, depending upon the phase
$\chi$. This is too big, compared to the observed value of
$V_{cb}\approx0.04\pm0.003$, by at least a factor of 3. We interpret
this failure as a {\it clue} to the presence of an antisymmetric
component in $M$, together with symmetrical ones
(so that $m_{ij}\neq m_{ji}$), which would modify the relevant mixing angle
to $\displaystyle{\sqrt{\frac{m_{i}}{m_{j}}}\sqrt{\frac{m_{ij}}{m_{ji}}}}$,
where $m_{i}$ and $m_{j}$ denote the respective eigenvalues. 

{\bf 3) The Need for a Contribution Proportional to $B$-$L$ :} The
success of the relations $m^{0}_{b}\approx m^{0}_{\tau}$, and
$m^{0}_{t}\approx m(\nu_{\tau})^{0}_{Dirac}$ (see Sec.\ref{Mass}), suggests
that the members of the third family get their masses primarily from
the VEV of a SU(4)-color singlet Higgs field that is independent of
$B$-$L$. This is in fact ensured if the Higgs is a 10 of
SO(10). However, the empirical observations of
$m^{0}_{s}\sim m^{0}_{\mu}/3$ and $m^{0}_{d}\sim 3m^{0}_{e}$ 
\cite{GeorgiJarsklog} clearly
call for a contribution proportional to $B$-$L$ as well. Further, one
can in fact argue that the suppression of $V_{cb}$ (in the
quark-sector) together with an enhancement of
$\theta^{osc}_{\nu_{\mu}\,\nu_{\tau}}$ (in the lepton sector) calls
for a contribution that is not only proportional to $B$-$L$, but 
also antisymmetric in the family space (as suggested above in item 
(\ref{e2})). We show below how both of these requirements can be met,
rather easily, in SO(10), even for a minimal Higgs system.

{\bf 4) Up-Down Asymmetry:} Finally, the up and
the down-sector mass matrices must not be proportional to each other, as
otherwise the CKM angles would all vanish. Note that the cubic couplings 
of a single $10_H$ will not serve the purpose in this regard.

Following Ref.\cite{BabuWilczekPati}, I now present a simple and predictive
mass-matrix, based on SO(10), that satisfies {\it all four}
requirements (\ref{e1}), (\ref{e2}), (\ref{e3}) and (\ref{e4}). The
interesting point is that one can obtain such a mass-matrix for the
fermions by utilizing only the minimal Higgs system, that is needed
anyway to break the gauge symmetry SO(10). It consists of the set :
\begin{eqnarray}
H_{minimal}\,=\,\{45_{H},\,16_{H},\,\overline{16}_{H},\,10_{H}\}\,.
\label{e14}
\end{eqnarray}
Of these, the VEV of $\langle45_{H}\rangle\sim M_{X}$ breaks SO(10)
into G(2213), and those of
$\langle16_{H}\rangle=\langle\overline{16}_{H}\rangle\sim M_{X}$ break
G(2213) to G(213), at the unification-scale $M_{X}$. Now G(213) breaks
at the electroweak scale by the VEV of $\langle10_{H}\rangle$ to
U$(1)_{em}\times$ SU$(3)^{c}$. 

One might have introduced large-dimensional tensorial multiplets of
SO(10) like $\overline{126}_{H}$ and 12$0_{H}$, both of which possess 
cubic level Yukawa couplings with the fermions. In particular, the
coupling $16_{i}16_{j}(120_{H})$ would give the desired
family-antisymmetric as well as ($B$-$L$)-dependent contribution. We do
not however introduce these multiplets in part because they do not
seem to arise in string solutions \cite{DienesRussell}, and in part also
because mass-splittings within such large-dimensional multiplets could 
give excessive threshold corrections to $\alpha_{3}(m_{z})$
(typically exceeding 20\%), rendering observed coupling
unification fortuitous. By contrast, the multiplets in the minimal set
(shown above) do arise in string solutions leading to
SO(10). Furthermore, the 
threshold corrections for the minimal set are found to be
 naturally small, and even to have the right sign,
to go with the observed coupling unification \cite{BabuWilczekPati} 
(see Appendix).

The question is: can the minimal set of Higgs multiplets 
(see Eq.(\ref{e14})) meet all the requirements listed
above? Now $10_{H}$ (even several 10`s) can not 
meet the requirements of antisymmetry and
$(B$-$L)$-dependence. Furthermore, a single $10_{H}$ cannot generate
CKM-mixings. This impasse disappears, however, as soon as one allows
 for not only
cubic, but also effective non-renormalizable quartic couplings of the
minimal set of Higgs fields with the fermions. These latter couplings
could of course well arise through exchanges of superheavy states
(e.g. those in
the string tower) involving renormalizable couplings, and/or through
quantum gravity. 

Allowing for such cubic and quartic couplings and adopting the
guideline (\ref{e1}) of hierarchical Yukawa couplings, as well as that
of economy, we are led to suggest the following effective lagrangian
for generating Dirac masses and mixings of the three families
\cite{BabuWilczekPati} (for a related but different pattern, involving a
non-minimal Higgs system, see Ref \cite{AlbrightBarr}).
\begin{eqnarray}
{\bf {\cal L}_{Yuk}}\,=\,h_{33}\,{\bf
16_{3}\,16_{3}\,10_{H}}\,+\,[\,h_{23}\,{\bf
16_{2}\,16_{3}\,10_{H}}\,+\,a_{23}\,{\bf
16_{2}\,16_{3}\,10_{H}\,45_{H}}/M\,\nonumber\\&&{\hspace{-12cm}}+\,g_{23}\,{\bf
16_{2}\,16_{3}\,16_{H}\,16_{H}}/M]\,+\,\{a_{12}\,{\bf
16_{1}\,16_{2}\,10_{H}\,45_{H}}/M\,\nonumber\\&&{\hspace{-10.5cm}}+\,g_{12}\,{\bf
16_{1}\,16_{2}\,16_{H}\,16_{H}}/M\}\,.
\label{e15}
\end{eqnarray}
Here, $M$ could plausibly be of order string scale. Note that a mass
matrix having essentially the form of Eq. (\ref{e11})  results if the
first term $h_{33}\langle10_{H}\rangle$ is dominant. This ensures
$m^{0}_{b}\approx m^{0}_{\tau}$ and $m^{0}_{t}\approx
m(\nu_{Dirac})^{0}$. Following the assumption of progressive hierarchy
(equivalently appropriate flavor symmetries
\footnote{Although no
  explicit string solution with the hierarchy in all the Yukawa couplings 
  in Eq.(\ref{e15}) - i.e. in $h_{ij}$, $a_{ij}$ and $g_{ij}$ - 
  exists as yet, one can postulate flavor symmetries of 
  the type alluded to (e.g. two abelian U(1) symmetries), which assign 
  flavor charges not only to the fermion families and the Higgs multiplets, 
  but also to a few (postulated) SM singlets that acquire VEVs of order 
  M$_X$. The flavor symmetry - allowed effective couplings such as 
  $16_2 16_3 10_H <S>/M$ would lead to $h_{23}\sim <S>/M \sim 1/10$. 
  One can verify that the full set of hierarchical couplings shown in 
  Eq.(\ref{e15}) can in fact arise in the presence of two such U(1) 
  symmetries. String theory 
  (at least) offers the scope (as indicated by the solutions of 
  Refs.\cite{FaraggiHalyo} and \cite{Antoniadis}) for providing a rationale 
  for the existence of such flavor symmetries, together with that of the 
  SM singlets. For example, there exist solutions with the 
  top Yukawa coupling being leading and others being hierarchical
  (as in Ref.\cite{FaraggiHalyo}).}
),we presume that $h_{23}\sim h_{33}/10$, while $h_{22}$ and $h_{11}$,
which are not shown, are assumed to be progressively much smaller than
$h_{23}$. Since 
$\langle45_{H}\rangle\sim\langle16_{H}\rangle\sim M_{X}$, while $M\sim
M_{st}\sim10M_{X}$, the terms $a_{23}\langle45_{H}\rangle/M$ and
$g_{23}\langle16_{H}\rangle/M$ can quite plausibly be of order
$h_{33}/10$, if $a_{23}\sim g_{23}\sim h_{33}$. By the assumption of
hierarchy, we presume that $a_{12}\ll a_{23}$, and $g_{12}\ll g_{23}$

It is interesting to observe the symmetry properties of the $a_{23}$
and $g_{23}$-terms.  Although $10_{H}\times45_{H}=10+120+320$, given
that $\langle45_{H}\rangle$ is along $B$-$L$, which is needed to
implement doublet-triplet  splitting (see Appendix), only 120 in the
decomposition contributes to the  mass-matrices.  This contribution
is, however, antisymmetric in the  family-index and, at the same time,
proportional to $B$-$L$. {\it Thus the  $a_{23}$ term fulfills the
requirements of both antisymmetry and ($B$-$L$)-dependence,
simultaneously
\footnote{The analog of $10_{H}\cdot45_{H}$ for the case
  of G(224) would be $\chi_{H}\equiv(2,2,1)_{H}\cdot(1,1,15)_{H}$.  
  Although in general, the
  coupling of $\chi_{H}$ to the fermions need not be antisymmetric, for
  a string-derived G(224), the multiplet (1,1,15$)_{H}$ is most likely
  to arise from an underlying 45 of SO(10) (rather than 210); in this
  case, the couplings of $\chi_{H}$ must be antisymmetric like that of
  $10_{H}\cdot45_{H}$.}
}. With only $h_{ij}$ and $a_{ij}$-terms,
however, the up and down quark  mass-matrices will be proportional to
each other, which would yield $V_{CKM} =1$.  This is remedied by the
$g_{ij}$ coupling, because, the $16_{H}$ can have a VEV not only along
its SM singlet component (transforming as
$\tilde{\overline{\nu}}_{R}$) which is of GUT-scale, but also along
its electroweak doublet component -- call it $16_{d}$ -- of the
electroweak  scale. The latter can arise by the  the mixing of
$16_{d}$ with the corresponding doublet (call it $10_{d}$) in the
$10_{H}$.  The MSSM doublet $H_{d}$, which is light, is then a mixture
of $10_{d}$ and $16_{d}$, while the orthogonal combination is
superheavy (see Appendix). Since $\langle16_{d}\rangle$ contributes
only to the down-flavor mass matrices, but not to the up-flavor, the
$g_{23}$ and $g_{12}$ couplings  generate non-trivial
CKM-mixings. {\it We thus see that the minimal  Higgs system 
(as shown in Eq.(\ref{e14})) satisfies
apriori all the qualitative requirements (1)-(4), including the
condition of $V_{CKM}\neq1$}. I now discuss that this  system works
well even quantitatively.

With these six effective Yukawa couplings, the Dirac mass matrices  of
quarks and leptons of the three families  at the unification scale
take the form :
\begin{eqnarray} 
U\,=\,\left(\begin{array}{ccc}{0} & {\epsilon'} & {0} \\
{-\,\epsilon'} & {0}  & {\epsilon\,+\,\sigma} \\ {0} &
{-\,\epsilon\,+\,\sigma} & {1}\end{array}\right)\,m_U,\,\,\,\,
D\,=\,\left(\begin{array}{ccc}{0} & {\epsilon'\,+\,\eta'} & {0} \\
{-\,\epsilon'\,+\,\eta'} & {0} & {\epsilon\,+\,\eta} \\ {0} &
{-\,\epsilon\,+\,\eta} & {1}\end{array}\right)\,m_D,  \nonumber \\[1.5em]
N=\left(\begin{array}{ccc}{0} & {-\,3\epsilon'} & {0} \\ {3\epsilon'}
& {0} & {-\,3\epsilon\,+\,\sigma} \\ {0} & {3\epsilon\,+\,\sigma} &
{1}\end{array}\right)\,m_U,\,  L=\left(\begin{array}{ccc}{0} &
{-\,3\epsilon'\,+\,\eta'} & {0} \\ {3\epsilon'\,+\,\eta'} & {0} &
{-\,3\epsilon\,+\,\eta} \\ {0} & {3\epsilon\,+\,\eta} &
{1}\end{array}\right)\,m_D. 
\label{e16}
\end{eqnarray} 
Here the matrices are multiplied by  left-handed fermion fields  from
the left and by anti--fermion fields from  the right.  $(U,D)$ stand
for the mass matrices of up and  down quarks, while $(N,L)$ are the
Dirac mass matrices  of the neutrinos and the charged leptons. The
entries $1,\epsilon$,and $\sigma$ arise respectively from the
$h_{33},a_{23}$ and $h_{23}$ terms in Eq. (\ref{e15}), while $\eta$
entering into $D$ and $L$ receives contributions from both  $g_{23}$
and $h_{23}$; thus $\eta\neq\sigma$. Similarly $\eta'$ and $\epsilon'$
arise from $g_{12}$ and $a_{12}$ terms respectively. Note the
quark-lepton correlations between $U$ and $N$ as well as $D$ and $L$,
and the up-down correlations between $U$ and $D$ as well as $N$ and
$L$. These correlations arise because of the symmetry property of
G(224). The relative factor of $-3$ between  quarks and leptons
involving the $\epsilon$ entry reflects the fact  that $\langle{\bf
45_{H}}\rangle\sim$ to(B-L), while the  antisymmetry in this entry
arises from the group structure of SO(10),  as explained
above$^4$. As we will see,
this $\epsilon$-entry helps to account for (a) the differences between
$m_{s}$ and $m_{\mu}$, (b) that between $m_{d}$ and $m_{e}$, and also,
(c) the suppression of $V_{cb}$ together with the enhancement of the
$\nu_{\mu}$-$\nu_{\tau}$ oscillation angle. 

The mass matrices in Eq.(\ref{e16}) contain 7 parameters
\footnote{
   Of these,  $m_{U}^{0}\approx m_{t}^{0}$ can in fact be estimated to
   within $20\%$  accuracy by either using the argument of radiative
   electroweak symmetry  breaking, or some promising string solutions
   (see e.g. Ref.\cite{FaraggiHalyo}).
}: $\epsilon$, $\sigma$, $\eta$, $m_{D}=h_{33}\,\langle10_{d}\rangle$,
$m_{U}=h_{33}\,\langle10_{U}\rangle$, $\eta'$ and $\epsilon'$. These
may be determined by using, for  example, the following input values:
$m_{t}^{phys}=174$ GeV, $m_{c}(m_{c})=1.37$ GeV, $m_{s}(1$
GeV$)=110$-$116$ MeV \cite{Gupta}, $m_{u}(1$ GeV) $\approx6$ MeV and
the observed masses of $e$, $\mu$ and $\tau$, which lead to (see
Ref.\cite{BabuWilczekPati}, for details):
\begin{eqnarray} 
\sigma\,\simeq\,\,0.110\,,\,\,\eta\simeq\,0.151\,,\,\,\epsilon\,\simeq\,-\,0.095\,,\,\,|\eta'|\approx4.4\times10^{-3}\,\,\,\mbox{and}\,\,\,\epsilon'\approx2\times10^{-4}\nonumber  
\end{eqnarray}
\begin{eqnarray}
m_{U}\,\simeq\,m_{t}(M_{U})\,\simeq\,(100\mbox{-}120)\,\mbox{GeV}\,,\,\,m_{D}\,\simeq\,m_{b}(M_{U})\,\simeq\,1.5\,\mbox{GeV}\,.
\label{e17} 
\end{eqnarray} 

Here, I will assume, only for the sake of simplicity, as in Ref. 
\cite{BabuWilczekPati}, that the parameters are real
\footnote{
   Babu and I have recently studied supersymmetric CP violation within 
   the G(224)/SO(10) 
   framework, by using precisely the fermion mass-matrices as in Eq.(\ref{e16}). 
   We have observed \cite{BabuJCP} that complexification of the parameters 
   can lead to observed CP violation, without upsetting in the least the 
   success of Ref.\cite{BabuWilczekPati} (i.e. of the fermion mass-matrices  
   of Eq.(\ref{e16})) in describing the masses and mixings of all fermions, 
   including neutrinos. Even with complexification the relative signs 
   and the approximate magnitudes of the real parts of the parameters must 
   be the same as in Eq.(\ref{e17}), to retain the success.
}. 
Note that in accord with our general
expectations  discussed above, each of the parameters $\sigma$, $\eta$
and $\epsilon$ are found to be of order 1/10,  as opposed to being
\footnote{
   This is one characteristic difference  between our work and
   that of Ref.\cite{AlbrightBarr}, where the (2,3)-element is even  
   bigger than the (3,3).
} $O(1)$ or $O(10^{-2})$, compared to the leading
(3,3)-element in Eq. (\ref{e16}). Having determined these parameters,
we are led to a total of five predictions involving only the quarks
(those for the leptons are listed separately) :
\begin{eqnarray} 
 m^{0}_{b}\,\approx\,m^{0}_{\tau}(1\,-\,8\epsilon^{2})\,;\,\,\,\,
\mbox{thus}\,\,\,\,m_{b}(m_{b})\,\simeq\,(4.6\mbox{-}4.9)\,\mbox{GeV}
\label{e18}
\end{eqnarray} 
\vspace*{-2em}
\begin{eqnarray} 
|V_{cb}|\,\simeq\,|\sigma\,-\,\eta|\,\approx\,
\left|\sqrt{m_{s}/m_{b}}\left|\frac{\eta\,+
\,\epsilon}{\eta\,-\,\epsilon}\right|^{1/2}\,-
\,\sqrt{m_{c}/m_{t}}\,\left|\frac{\sigma\,+\,\epsilon}{\sigma\,-
\,\epsilon}\right|^{1/2}\right|\,\simeq\,0.045 
\label{e19} 
\end{eqnarray} 
\vspace*{-2em}
\begin{eqnarray} 
m_{d}\,(1 \mbox{GeV})\,\simeq\,8\,\mbox{MeV}
\label{e20} 
\end{eqnarray} 
\vspace*{-2em}
\begin{eqnarray} 
\theta_{C}\,\simeq\,\left|\sqrt{m_{d}/m_{s}}\,-
 \,e^{i\phi}\sqrt{m_{u}/m_{c}}\right|
\label{e21} 
\end{eqnarray} 
\vspace*{-2em}
\begin{eqnarray} 
|V_{ub}/V_{cb}|\,\simeq\,\sqrt{m_{u}/m_{c}}\,\simeq\,0.07\,.
\label{e22} 
\end{eqnarray} 
In making these predictions, we have extrapolated the GUT-scale values
down to low energies using $\alpha_{3}(m_{Z})=0.118$, a SUSY threshold
of 500 GeV and $\tan\beta=5$. The results depend  weakly on these
choices, assuming  $\tan\beta\approx2$-30. Further, the Dirac masses and
mixings of the neutrinos and the mixings of the  charged  leptons also
get determined. We obtain :
\begin{eqnarray} 
m_{\nu_{\tau}}^{D}(M_{U})\,\approx\,100\mbox{-}120\,
\mbox{GeV};\,\,m_{\nu_{\mu}}^{D}(M_{U})\,\simeq\,8\,\mbox{GeV},
\label{e23} 
\end{eqnarray} 
\vspace*{-2em}
\begin{eqnarray} 
\theta_{\mu\tau}^{\ell}\,\approx\,-\,3\epsilon\,+\,\eta\,
\approx\,\sqrt{m_{\mu}/m_{\tau}}\,\left|\frac{-\,3\epsilon\,+
\,\eta}{3\epsilon\,+\,\eta}\right|^{1/2}\,\simeq\,0.437
 \label{e24} 
\end{eqnarray} 
\vspace*{-2em}
\begin{eqnarray} 
m_{\nu_{e}}^{D}\,\simeq\,[\,9\epsilon^{'2}/(9\epsilon^{2}\,-\,\sigma^{2})]\,m_{U}\,\simeq\,0.4\,\mbox{MeV}
\label{e25} 
\end{eqnarray} 
\vspace*{-2em}
\begin{eqnarray} 
\theta_{e\mu}^\ell\,\simeq\,\left|\frac{\eta'\,-\,3\epsilon'}{\eta'\,+\,3\epsilon'}\right|^{1/2}\,\sqrt{m_{e}/m_{\mu}}\,\simeq\,0.85\,\sqrt{m_{e}/m_{\mu}}\,\simeq\,0.06
\label{e26}
\end{eqnarray} 
\vspace*{-1.5em}
\begin{eqnarray} 
\theta_{e\tau}^\ell\,\simeq\,\frac{1}{0.85}\,\sqrt{m_{e}/m_{\tau}}\,(m_{\mu}/m_{\tau})\,\simeq\,0.0012\,.
\label{e27} 
\end{eqnarray} 
In evaluating $\theta_{e\mu}^\ell$, we have assumed $\epsilon'$ and
$\eta'$ to be relatively positive.

Given the bizarre pattern of quark and lepton masses and mixings, it
seems remarkable that the simple pattern of fermion mass-matrices,
motivated by the group theory of G(224)/SO(10), gives an overall fit
to  all of them (Eqs.(\ref{e18}) through (\ref{e22})) which is good 
to within $10\%$.  This includes the two
successful predictions on $m_{b}$ and $V_{cb}$ (Eqs.(\ref{e18}) and
(\ref{e19})). Note that in supersymmetric unified theories, the
``observed'' value of  $m_{b}(m_{b})$ and renormalization-group
studies suggest that, for a wide  range of the parameter $\tan\beta$,
$m_{b}^{0}$ should in fact be  about 10-20$\%$ {\it lower} than
$m_{\tau}^{0}$ \cite{PierceBabuKolda}.  This is  neatly explained by the
relation: $m_{b}^{0}\approx m_{\tau}^{0}(1 - 8\epsilon^{2})$
(Eq. (\ref{e18})), where exact equality holds in the limit
$\epsilon\rightarrow0$ (due to SU(4)-color), while the decrease of
$m^{0}_{b}$ compared to $m^{0}_{\tau}$ by $8\epsilon^{2}\sim10\%$ is
precisely because the off-diagonal $\epsilon$-entry is proportional to
$B$-$L$ (see Eq. (\ref{e16})).    
 
Specially intriguing is the result on $V_{cb}\approx0.045$ which
compares well with the observed value of $\simeq0.04$.  The
suppression  of $V_{cb}$, compared to the value of $0.17 \pm 0.06$
obtained from Eq.  (\ref{e13}), is now possible because the mass
matrices (Eq. (\ref{e16})) contain an antisymmetric component
$\propto\epsilon$. That corrects the square-root  formula
$\theta_{sb}=\sqrt{m_{s}/m_{b}}$ (appropriate for symmetric matrices,
see Eq. (\ref{e11})) by the asymmetry factor
$|(\eta+\epsilon)/(\eta-\epsilon)|^{1/2}$ (see Eq. (19)), and
similarly for the angle $\theta_{ct}$. This factor suppresses $V_{cb}$
if $\eta$ and $\epsilon$ have opposite signs. The interesting point is
that, {\it the same feature necessarily enhances the corresponding
mixing  angle $\theta_{\mu\tau}^{\ell}$ in the leptonic sector}, since
the  asymmetry factor in this case is given by
$[(-3\epsilon+\eta)/(3\epsilon+\eta)]^{1/2}$ (see Eq. (24)). This
enhancement of $\theta_{\mu\tau}^\ell$ helps to account for the nearly
maximal oscillation angle observed at SuperK (as discussed
below). This intriguing correlation between the mixing angles in the
quark versus leptonic sectors -- {\it that is suppression of one
implying enhancement of the other} -- has become possible only because
of the $\epsilon$-contribution, which is simultaneously antisymmetric
and is proportional to $B$-$L$. That in turn becomes possible because
of the group-property of SO(10) or a string-derived G(224)$^{4}$. 

Taking stock, we see an overwhelming set of facts in favor of
$B$-$L$ and  in fact for the full SU(4)-color-symmetry.    These
include: (i) the suppression of $V_{cb}$, together with the
enhancement of $\theta_{\mu\tau}^{\ell}$, just mentioned above, (ii)
the successful relation  $m_{b}^{0}\approx
m_{\tau}^{0}(1-8\epsilon^{2})$, (iii) the usefulness  again of the
SU(4)-color-relation $m(\nu_{Dirac}^{\tau})^{0}\approx m_{t}^{0}$ in
accounting for $m(\nu_{L}^{\tau})$( see Sec. 4 ),  and (iv) the
agreement of the relation
$|m_{s}^{0}/m_{\mu}^{0}|=|(\epsilon^{2}-\eta^{2})/(9\epsilon^{2}-\eta^{2})|$
with the data, in that the ratio   is naturally {\it less than} 1, if
$\eta\sim\epsilon$.    The presence of  $9\epsilon^2$ in the
denominator is because the off-diagonal entry is  proportional to
B-L. Finally, the need for  ($B$-$L$)- as a local symmetry, to
implement baryogenesis, has been noted in Sec.1.

Turning to neutrino masses, while all the entries in the Dirac mass
matrix $N$ are now fixed, to obtain the parameters for the light
neutrinos, one needs to specify those of the Majorana mass matrix of
the RH neutrinos ($\nu^{e,\mu,\tau}_{R}$). Guided by economy and the
assumption of hierarchy, we consider the following pattern :
\begin{eqnarray} 
M_{\nu}^{R} = \left(\begin{array}{ccc}{x} & {0} & {z} \\ {0} & {0} &
{y} \\ {z} & {y} & {1}\end{array}\right)\,M_{R}\,. 
\label{e28}
\end{eqnarray} 

 As discussed in Sec.\ref{Mass}, the  magnitude of
$M_{R}\approx(5\mbox{-}15)\times10^{14}$ GeV can quite plausibly be
justified in the context of supersymmetric unificaton\footnote{This
estimate for $M_{R}$ is retained even if one allows for
$\nu_{\mu}\mbox{-}\nu_{\tau}$ mixing (see Ref. \cite{BabuWilczekPati}).} 
(e.g. by using $M\approx M_{st}\approx4\times10^{17}$ GeV in Eq. (\ref{e8})).
 To the same extent,  the
magnitude of $m(\nu_{\tau})\approx(1/10\mbox{-}1/30)$ eV, which is
consistent with the SuperK value, can also be anticipated. Thus there
are effectively three new parameters: $x$, $y$, and $z$. Since there
are six observables for the three light neutrinos,
 one can expect three predictions.
These may be taken to be $\theta_{\nu_{\mu}\nu_{\tau}}^{osc}$,
$m_{\nu_{\tau}}$ (see Eq. (\ref{e10})), and for example
$\theta_{\nu_{e}\nu_{\mu}}^{osc}$.

Assuming successively hierarchical entries as for the Dirac mass matrices,
we presume that $|y|\sim1/10,
|z|\leq|y|/10$ and $|x|\leq z^{2}$. Now given that $m(\nu_{\tau})\sim1/20$
eV (as estimated in Eq. (\ref{e10})),
the MSW solution for the solar neutrino puzzle \cite{MSW} suggests
that $m(\nu_{\mu})/m(\nu_{\tau})\approx1/10\mbox{-}1/30$. The latter
in turn yields :  $|y|\approx(1/18\mbox{\,\,to\,\,}1/23.6)$, with $y$
having the same sign as $\epsilon$ (see Eq. (\ref{e17})). This
solution for y  obtains only by assuming that $y$ is $O(1/10)$ rather than
$O(1)$. Combining now with the mixing in the $\mu$-$\tau$ sector
determined above (see Eq. (\ref{e24})), one can then determine the
$\nu_{\mu}\mbox{-}\nu_{\tau}$ oscillation angle. The two predictions
of the model for the neutrino-system are then :
\begin{eqnarray}
m(\nu_{\tau})\,\approx\,(1/10\,\mbox{-}\,1/30)\,\mbox{eV}
\label{e29n}
\end{eqnarray}
\vspace*{-2em}
\begin{eqnarray}
\theta_{\nu_{\mu}\nu_{\tau}}^{osc}\,\simeq\,\theta_{\mu\tau}^{\ell} - 
 \theta_{\mu\tau}^{\nu}\,\simeq
\,\left(0.437\,+\,\sqrt{\frac{m_{\nu_{2}}}{m_{\nu_{3}}}}\,\right)\, .
\label{e29}
\end{eqnarray}
\vspace*{-2em}
\begin{eqnarray}
\mbox{Thus,}\,\,\,{\sin}^{2}\,2\theta_{\nu_{\mu}\nu_{\tau}}^{osc}=
(0.96,0.91,0.86,0.83,0.81)\,\,\,\,\\
\mbox{for}\,\,\,\, m_{\nu_{2}}/m_{\nu_{3}}=(1/10,1/15,1/20,1/25,1/30)\,.
\label{e30}
\end{eqnarray}
Both of these predictions are extremely successful.

Note the interesting point that the MSW solution, together with the
requirement that $|y|$ should have a natural  hierarchical value (as
mentioned above),
lead to $y$ having the same sign as $\epsilon$; that (it turns
out) implies that the two contributions in Eq.(\ref{e29}) must {\it
add} rather than subtract, leading to an {\it almost maximal
oscillation angle\,}\cite{BabuWilczekPati}. 
The other factor contributing to the
enhancement of $\theta_{\nu_{\mu}\nu_{\tau}}^{osc}$ is, of course,
also the asymmetry-ratio which increases $|\theta_{\mu\tau}^{\ell}|$
from 0.25 to 0.437 (see Eq. (\ref{e24})). We see that  one can derive
rather plausibly a large $\nu_{\mu}\mbox{-}\nu_{\tau}$ oscillation
angle $\sin^{2}\,2\theta_{\nu_{\mu}\nu_{\tau}}^{osc}\geq0.8$, together
with an understanding of  hierarchical masses and mixings of the
quarks and the charged leptons,  while maintaining a  large hierarchy
in the seesaw derived neutrino masses
($m_{\nu_{2}}/m_{\nu_{3}}=1/10\mbox{-}1/30$), all within a unified
framework  including both quarks and leptons.  In the example
exhibited  here, the mixing angles for the mass eigenstates of neither
the neutrinos  nor the charged leptons are really large, in that
$\theta_{\mu\tau}^{\ell}\simeq0.437\simeq23^{\circ}$ and
$\theta_{\mu\tau}^{\nu}\simeq(0.18\mbox{-}0.31)\approx(10\mbox{-}18)^{\circ}$,
{\it yet the oscillation angle obtained by combining the two is
near-maximal.}  This contrasts with most works in the literature in which a
large  oscillation angle is obtained either entirely from the neutrino
sector  (with nearly  degenerate neutrinos) or almost entirely from
the charged lepton sector.

While $M_{R}\approx(5\mbox{-}15)\times10^{14}$ GeV and $y\approx-1/20$
are better determined, the parameters $x$ and $z$ can not be obtained
reliably at present because very little is known
about observables involving $\nu_{e}$. Taking, for concreteness,
$m_{\nu_{e}}\approx(10^{-5}\mbox{-}10^{-4}$ (1 to few)) eV and
$\theta^{osc}_{e\tau}\approx\theta^{\ell}_{e\tau}-\theta^{\nu}_{e\tau}\approx10^{-3}\pm0.03$
as inputs, we obtain : $z\sim(1$-$5)\times10^{-3}$ and $x\sim($1 to
few)$(10^{-6}\mbox{-}10^{-5})$, in accord with the guidelines of
$|z|\sim|y|/10$ and $|x|\sim z^{2}$. This in turn yields :
$\theta^{osc}_{e\mu}\approx\theta^{\ell}_{e\mu}-\theta^{\nu}_{e\mu}\approx0.06\pm0.015$.
Note that the mass of $m_{\nu_{\mu}}\sim3\times10^{-3}$ eV, that follows
from a
natural hierarchical value for $y\sim-(1/20)$, and $\theta_{e\mu}$ as
above, go well with the small angle MSW explanation\footnote{Although
the small angle MSW solution appears to be more generic within the
approach outlined above, we have found that the large angle solution
can still plausibly emerge in a limited region of parameter space,
without affecting our results on fermion masses.} of the solar
neutrinos puzzle. 

It is worthnoting that although the superheavy Majorana masses of the RH
neutrinos cannot be
observed directly, they can be of cosmological significance.  The
pattern given above and the arguments given in Sec.\ref{Need} and in this
section suggests that
$M(\nu_{R}^{\tau})\approx(5\mbox{-}15)\times10^{14}$  GeV,
$M(\nu_{R}^{\mu})\approx(1\mbox{-}4)\times10^{12}$ GeV (for
$x\approx1/20$); and $M(\nu_{R}^{e})\sim(1/2\mbox{-}10)\times10^9$ GeV
(for $x\sim(1/2\mbox{-}10)10^{-6}>z^2$).    A mass of
$\nu_{R}^{e}\sim10^{9}$ GeV is of the   right magnitude for producing
$\nu_{R}^{e}$ following reheating and inducing lepton asymmetry   in
$\nu_{R}^{e}$ decay into $H^{0}+\nu_{L}^{i}$, that is subsequently
converted into baryon asymmetry by the electroweak sphalerons
\cite{KuzminRubakov,LeptoB}.

In summary, we have proposed an economical and predictive pattern for
the Dirac  mass matrices, within the SO(10)/G(224)-framework, which is
remarkably successful in describing the observed masses and  mixings
of {\it all} the quarks and charged leptons. It leads to five
predictions for just the quark- system, all of which agree with
observation to
within 10\%.  The same  pattern, supplemented with a similar structure
for the Majorana mass matrix, accounts for both the large
$\nu_{\mu}$-$\nu_{\tau}$ oscillation angle and a mass of
$\nu_{\tau}\sim1/20$ eV, suggested by the SuperK data. 
Given this degree of
success, it makes good sense to study proton decay concretely within
this SO(10)/G(224)-framework. The results of this study 
\cite{BabuWilczekPati} are presented in the next section.

Before turning to proton decay, it is worth noting that much of our
discussion  of fermion masses and mixings, including those of the
neutrinos, is  essentially unaltered if we go to the limit
$\epsilon'\rightarrow0$ of  Eq. (28).  This limit clearly involves: 
\begin{eqnarray} 
m_{u}\,=\,0\,,\,\,\,\,\theta_{C}\,\simeq\,\sqrt{m_{d}/m_{s}}\,,\,\,\,\,m_{\nu_{e}}\,=\,0\,,\,\,\,\,\theta_{e\mu}^{\nu}\,=\,\theta_{e\tau}^{\nu}\,=\,0\,.  
\nonumber 
\end{eqnarray}
\begin{eqnarray} 
|V_{ub}|\,\simeq\,\sqrt{\frac{\eta\,-\,\epsilon}{\eta\,+\,\epsilon}}\,\sqrt{m_{d}/m_{b}}\,(m_{s}/m_{b})\,\simeq\,(2.1)(0.039)(0.023)\,\simeq\,0.0019 
\label{e32}
\end{eqnarray} 
All other predictions  remain unaltered.  Now, among the observed
quantities in the list above, $\theta_C\simeq\sqrt{m_{d}/m_{s}}$ is a
good result. Considering that $m_{u}/m_{t}\approx10^{-5}$, $m_{u}=0$
is  also a pretty good result.  There are of course plausible small
corrections  which could arise through Planck scale physics; these
could  induce a small value for $m_{u}$ through the (1,1)-entry
$\delta\approx10^{-5}$.    For considerations of proton decay, it is
worth distinguishing between these two {\it extreme} variants
which we will refer to as cases I and II respectively.

\begin{eqnarray} 
\mbox{Case I
:}&&\epsilon'\,\approx\,2\,\times\,10^{-4}\,,\,\,\,\delta\,=\,0
\nonumber \\[1em]
\mbox{Case II :}&&\delta\,\approx\,10^{-5}\,,\,\,\,\epsilon'\,= 0\,. 
\label{e33}
\end{eqnarray} 
It is worth noting that the observed value of $|V_{ub}|\approx 0.003$ 
favors a non-zero value of $\epsilon'(\approx (1-2)\times 10^{-4})$. 
Thus, in reality, $\epsilon'$ may not be zero, but it may lie in 
between the two extreme values listed above. 
In this case, the predicted proton lifetime for the 
standard $d=5$ operators would be intermediate between those for the 
two cases, presented in Sec.\ref{Expectations}.

\section{Expectations for Proton Decay in Supersymmetric Unified 
         Theories} 
\label{Expectations}
\subsection{Preliminaries}\label{Preliminaries}
Turning to the main purpose of this talk, I
present now the reason why the unification framework based on SUSY
SO(10) or G(224), together with the understanding of fermion masses
and mixings discussed above, strongly suggest that proton decay should
be imminent.

Recall that supersymmetric unified theories (GUTs) introduce two new
features to proton decay : (i) First, by raising $M_{X}$ to a higher
value of about $2\times10^{16}$ GeV (contrast with the non-supersymmetric 
case of nearly $3\times 10^{14}$ GeV), they strongly suppress the
gauge-boson-mediated $d=6$ proton decay operators, for which
$e^{+}\pi^{0}$ would have been the dominant mode (for this case, one
typically obtains : $\Gamma^{-1}(p\rightarrow
e^{+}\pi^{0})|_{d=6}\approx10^{35.3\pm1.5}$ yrs). (ii) Second, they
generate $d=5$ proton decay operators \cite{Sakai} of the form
$Q_{i}Q_{j}Q_{k}Q_{l}/M$ in the superpotential, through the exchange
of color triplet Higginos, which are the GUT partners of the standard
Higgs(ino) doublets, such as those in the $5+\overline{5}$ of SU(5) or
the 10 of SO(10). Assuming that a suitable doublet-triplet splitting
mechanism provides heavy GUT-scale masses to these color triplets 
and at the same time 
light masses to the doublets, these ``standard'' $d=5$ operators,
suppressed by just one power of the heavy mass and the small Yukawa
couplings, are found to provide the dominant mechanism for proton
decay in supersymmetric GUT \cite{DimopRaby,Nath,Hisano,BabuBarr}.

Now, owing to (a) Bose symmetry of the superfields in $QQQL/M$, (b)
color antisymmetry, and especially (c) the hierarchical Yukawa
couplings of the Higgs doublets, it turns out that these standard
$d=5$ operators lead to dominant $\overline{\nu}K^{+}$ and comparable
$\overline{\nu}\pi^{+}$ modes, but in all cases to highly suppressed
$e^{+}\pi^{0}$, $e^{+}K^{0}$ and even $\mu^{+}K^{0}$ modes. For
instance, for minimal SUSY SU(5), one obtains (with $\tan\beta\leq20$,
say) : 
\begin{eqnarray}
[\,\Gamma(\mu^{+}K^{0})/\Gamma(\overline{\nu}K^{+})\,]^{SU(5)}_{std}\,\sim\,[m_{u}/(m_{c}\,\sin^2\theta_c)]^2\,R\,\approx\,10^{-3}\,,
\label{e34}
\end{eqnarray}
where $R\approx0.1$ is the ratio of the relevant $|$matrix
element$|^{2}\times$(phase space), for the two modes.

It was recently pointed out that in SUSY unified theories based on
SO(10) or G(224), which assign heavy Majorana masses to the RH
neutrinos, there exists a new set of color triplets and thereby very
likely a {\it new source} of $d=5$ proton decay operators
\cite{BPW1}. For instance, in the context of the minimal set of Higgs
multiplets\footnote{The origin of the new $d=5$ operators in the
context of other Higgs multiplets, in particular in the cases where
$126_{H}$ and $\overline{126}_{H}$ are used to break $B$-$L$, has
been discussed in Ref.\cite{BPW1}.}
$\{45_{H},16_{H},\overline{16}_{H}$ and $10_{H}\}$ 
(see Sec.\ref{Understanding}), these
new $d=5$ operators arise by combining three effective couplings
introduced before :-- i.e., (a) the couplings
$f_{ij}16_{i}16_{j}\overline{16}_{H}\overline{16}_{H}/M$ (see
Eq.(\ref{e7})) that are required to assign Majorana masses to the RH
neutrinos, (b) the couplings $g_{ij}16_{i}16_{j}{16}_{H}{16}_{H}/M$,
which are needed to generate non-trivial CKM mixings (see
Eq.(\ref{e15})), and (c) the mass term
$M_{16}16_{H}\overline{16}_{H}$. 
For the $f_{ij}$ couplings, there are two possible SO(10)-contractions  
(leading to a 45 or a 1) for the pair $16_i\overline{16}_H$, both 
of which contribute to the Majorana masses of the RH neutrinos, 
but only the non-singlet contraction (leading to 45), would contribute 
to d=5 proton decay operator. In the presence of non-perturbative 
quantum gravity, one would in general expect the two contractions to have 
comparable strength. Furthermore, the couplings of 45's lying in the 
string-tower or possibly below the string-scale, and likewise of 
singlets, to the $16_i\cdot\overline{16}_H$-pair, would respectively 
generate the two contractions. It thus seems most likely that both 
contractions are present, having comparable strength. Allowing for 
a difference between the relevant projection factors for $\nu_R$ 
masses versus proton decay, and also for the fact that both contractions 
contribute to the former, but only the non-singlet one (i.e. 45) to 
the latter, we would set the relevant $f_{ij}$ coupling for proton 
decay to be $(f_{ij})_p\equiv(f_{ij})_\nu\cdot K$, where $(f_{ij})_\nu$ 
defined in Sec.\ref{Mass} directly yields $\nu_R$ - masses 
(see Eq.(\ref{e8})); and K is a relative factor of order unity. As a 
plausible range, we will take $K\approx 1/3$ to 2 (say). In the 
presence of the non-singlet contraction,  the
color-triplet Higginos in $\overline{16}_{H}$ and $16_{H}$ of mass
$M_{16}$ can be exchanged between $\tilde{q}_{i}q_{j}$ and
$\tilde{q}_{k}q_{l}$-pairs (correspondingly, for G(224), the color triplets 
would arise from $(1,2,4)_H$ and $(1, 2, \overline{4})_H$). 
This exchange generates a new set of $d=5$ 
operators in the superpotential of the form
\begin{eqnarray}
W_{new}\,\propto\,(f_{ij})_\nu\,g_{kl}K\,(16_{i}\,16_{j})\,(16_{k}\,16_{l})\,\langle\overline{16}_{H}\rangle\,\langle{16}_{H}\rangle/M^2\,\times (1/M_{16}),
\label{e35}
\end{eqnarray}
which induce proton decay. Note that these operators depend, through
the couplings $f_{ij}$ and $g_{kl}$, both on the Majorana and on the
Dirac masses of the respective fermions. {\it This is why within SUSY
SO(10) or G(224), proton decay gets intimately linked to the masses
and mixings of all fermions, including neutrinos.}

\subsection{Framework for Calculating Proton Decay Rate}
\label{Framework}

To establish notations, consider the case of minimal SUSY SU(5) and, as 
an example, the process
$\tilde{c}\tilde{d}\rightarrow{\bar{s}}{\bar{\nu}_{\mu}}$, which
induces $p\rightarrow\overline{\nu}_{\mu}K^{+}$. Let the strength of
the corresponding $d=5$ operator, multiplied by the product of the CKM
mixing elements entering into wino-exchange vertices, 
(which in this case is $\sin\theta_{C}\cos\theta_{C})$  
be denoted by $\hat{A}$. Thus (putting $\cos\theta_{C} =1$), one obtains:  
\begin{eqnarray} 
{\hspace{-0.3cm}}\hat{A}_{\tilde{c}\tilde{d}}(SU(5))\,=\,(h_{22}^{u}\,h_{12}^{d}/M_{H_{C}})\,\sin\theta_{c}\,\simeq\,(m_{c}m_{s}\,\sin^{2}\theta_{C}/v_{u}^{2})\,(\tan\beta/M_{H_{C}})\nonumber\\&&{\hspace{-13.3cm}}\simeq\,(1.9\times10^{-8})\,(\tan\beta/M_{H_C})\,\approx\,(2\times10^{-24}\,\mbox{GeV$^{-1}$})\,(\tan\beta/2)\,(2\times10^{16}\,\mbox{GeV}/M_{H_{C}})\,, 
\label{e36} 
\end{eqnarray}
where $\tan\beta\equiv{v}_{u}/v_{d}$, and we have put $v_{u}=174$ GeV
and the fermion masses extrapolated to the unification-scale --
i.e. $m_{c}\simeq300$ MeV and $m_{s}\simeq40$ MeV.  The amplitude for the associated four-fermion process $dus\rightarrow\overline{\nu}_{\mu}$ is given by: 
\begin{eqnarray} 
A_5(dus\,\rightarrow\,\overline{\nu}_{\mu})\,=\,\hat{A}_{\tilde{c}\tilde{d}}\,\times\,(2f)
\label{e37} 
\end{eqnarray} 
where $f$ is the loop-factor associated with wino-dressing. Assuming 
$m_{\tilde{w}}\ll{m}_{\tilde{q}}\sim{m}_{\tilde{l}}$, one gets:
$f\simeq(m_{\tilde{w}}/m^{2}_{\tilde{q}})(\alpha_{2}/4\pi)$.  
Using the amplitude for $(du)(s\nu_\ell$), as in Eq. (\ref{e37}), 
($\ell=\mu$ or $\tau$), one then obtains 
\cite{Nath,Hisano,BabuBarr,BabuWilczekPati}:  
\begin{eqnarray} 
\Gamma^{-1}(p\,\rightarrow\,\overline{\nu}_{\tau}K^+)\,\approx\,(0.6\times10^{31})\,\mbox{yrs}\times\nonumber\\&&\hspace{-7cm}\left(\frac{0.67}{{A}_{S}}\right)^2\,\left[\frac{0.014\,\mbox{GeV}^3}{\beta_{H}}\right]^{2}\,\left[\frac{(1/6)}{(m_{\tilde{W}}/m_{\tilde{q}})}\right]^{2}\left[\frac{m_{\tilde{q}}}{1.2\,\mbox{TeV}}\right]^2\,\left[\frac{2\times10^{-24}\,\mbox{GeV}^{-1}}{\hat{A}(\overline{\nu})}\right]^2\,. 
\label{e38}
\end{eqnarray} 
Here $\beta_{H}$ denotes the hadronic matrix element defined by
$\beta_{H}
u_{L}(\vec{k})\equiv\epsilon_{\alpha\beta\gamma}\langle0|(d_{L}^{\alpha}u_{L}^{\beta})u_{L}^{\gamma}|p,\vec{k}\rangle$.
While the range $\beta_{H}=(0.003\mbox{-}0.03)$ GeV$^{3}$ has been
used in the  past \cite{Hisano}, given that one  lattice calculation
yields $\beta_{H}=(5.6\pm0.5)\times10^{-3}$ GeV$^3$ \cite{JLQCD99},  
and a recent improved calculation yields $\beta_H\approx 0.014$GeV$^3$
\cite{Aoki} (whose systematic errors that may arise from scaling 
violations and quenching are hard to estimate \cite{Aoki}), 
we will take as a conservative, but plausible, range 
for $\beta_H$ to be given by $(0.014$GeV$^3)(1/2 - 2)$. [Compare this with 
the range for $\beta_H = (0.006$GeV$^3)(1/2 - 2)$ as used in 
Ref.\cite{BabuWilczekPati}].
Here, $A_{S}\approx0.67$ stands for the short distance renormalization
factor of the $d=5$ operator. Note that the 
familiar factors that appear in the expression for proton 
lifetime 
-- i.e., $M_{H_{C}}$, ($1+y_{tc}$) representing the interference between
the 
$\tilde{t}$ and $\tilde{c}$ contributions, and $\tan\beta$ (see
e.g. Ref.\cite{Hisano} and discussion in the Appendix of 
Ref.\cite{BabuWilczekPati}) -- are all effectively 
contained in $\hat{A}(\overline{\nu})$.
In Ref.\cite{BabuWilczekPati}, guided by the demand of naturalness 
(i.e. absence of excessive fine tunning) in obtaining the Higgs boson mass, 
squark masses were assumed to lie in the range of 
1 TeV$(1/\sqrt{2} - \sqrt{2})$, 
so that $m_{\tilde{q}}\lesssim 1.4$TeV. Recent work, based on the notion 
of focus point supersymmetry however suggests that squarks may be considerably 
heavier without conflicting with the demands of naturalness \cite{Feng+}. In 
the interest of obtaining a conservative upper limit on proton lifetime, 
we will therefore allow squark masses to be as heavy as about 2.5 TeV 
and as light as perhaps 600 GeV.
\footnote{
  We remark that if the recently reported (g-2) - anomaly for the muon 
  \cite{g2} is attributed to supersymmetry \cite{SUSYg2}, one would 
  need to have extremely light s-fermions (i.e. 
  $m_{\tilde{l}}\approx 200 - 400$ GeV (say) and correspondingly (for 
  promising mechanisms of SUSY-breaking) $m_{\tilde{q}}\lesssim 300 - 600 $
  TeV (say)), and simultaneously large or very large $\tan\beta (\approx 25 
  - 50)$. However, not worring about grand unification, such light 
  s-fermions, together with large or very large $\tan\beta$ would 
  typically be in gross conflict with the limits on the edm's of the neutron 
  and the electron, unless on can explain naturally the occurence of 
  minuscule phases $(\lesssim 1/300$ to $1/1000$) and/or large cancellation. 
  Thus, if the $(g-2)_{\mu}$ - anomaly turns out to be real, it may quite 
  possibly need a non-supersymmetric explanation, in accord with the 
  edm-constraints which ordinarily seem to suggest that squarks are (at least) 
  moderately heavy ($m_{\tilde{q}}\gtrsim 0.6 - 1$ TeV, say), and $\tan\beta$ 
  is not too large ($\lesssim 3$ to $10$, say).
  We mention in passing that the extra vector - like matter - specially a 
  $16 + \overline{16}$ of SO(10) -  as proposed in the so-called extended 
  supersymmetric standard model (ESSM) \cite{BabuJi,BabuPatiStrem},  
  with the heavy lepton mass being of order (150-200)
  hundred GeV, can provide such an explanation 
  \cite{BabuJCP2}. Motivations for the case of ESSM, based on the need for 
  (a)  
  removing the mismatch between MSSM and string unification scales, and 
  (b) dilaton-stabilization, have been noted in Ref.\cite{BabuJi}. Since ESSM 
  is an interesting and viable variant of MSSM, and would have important 
  implications for proton decay, we will present the results for expected 
  proton decay rates for the cases of both MSSM and ESSM in the discussion 
  to follow.
}
 
Allowing for plausible and rather generous uncertainties in the matrix 
element and the spectrum we take: 
\begin{eqnarray} 
\beta_{H}\,=\,(0.014\,\mbox{GeV}^3)\,(1/2\,\mbox{-}\,2) 
\nonumber 
\end{eqnarray}
\begin{eqnarray} 
m_{\tilde{w}}/m_{\tilde{q}}\,=\,1/6\,(1/2\,\mbox{-}\,2)\,,\,\,\,\,{\rm and}\,\,\,\,m_{\tilde{q}}\,\approx\,m_{\tilde{\ell}}\,\approx\,1.2\,\mbox{TeV}\,(1/2\,\mbox{-}\,2)\,. 
\label{e39}
\end{eqnarray}
Using Eqs.(\ref{e38}-\ref{e39}), we get: 
\begin{eqnarray} 
\Gamma^{-1}(p\,\rightarrow\,\overline{\nu}_{\tau}K^{+})\,\approx\,(0.6\times10^{31}\,\mbox{yrs})\,[\,2\times10^{-24}\,\mbox{GeV}^{-1}/\hat{A}(\overline{\nu}_\ell)\,]^{2}\,\{64\,\mbox{-}\,1/64\,\}\,.
\label{e40} 
\end{eqnarray} 
Note that the curly bracket would acquire its upper-end value of 64, which 
would serve towards maximizing proton lifetime, only provided all the 
uncertainties in Eq.(\ref{e40}) are stretched to the extreme so that 
$\beta_H=0.007$ GeV$^3$, $m_{\tilde{W}}/m_{\tilde{q}}\approx 1/12$ and 
$m_{\tilde{q}}\approx 2.4$ TeV. 
This relation, as well as Eq. (\ref{e38}) are general, depending only
on $\hat{A}(\overline{\nu}_{\ell})$ and on the range of parameters
given in  Eq. (\ref{e39}). They can thus be used for both SU(5) and
SO(10).

The experimental lower limit on the inverse rate for the  
$\bar{\nu}K^{+}$ modes is given by \cite{SKlimit},  
\begin{eqnarray} 
[\sum_{\ell}\,\Gamma(p\,\rightarrow\,\overline{\nu}_{\ell}K^{+})]^{-1}_{expt} 
	\geq 1.6\times 10^{33}\,\mbox{yrs}\,. 
\label{e41}
\end{eqnarray} 
Allowing for all the uncertainties to stretch in the same  direction
(in this case, the curly bracket = 64), and assuming  that just one
neutrino flavor (e.g. $\nu_{\mu}$ for SU(5)) dominates, the   observed
limit (Eq.(\ref{e41})) provides an upper bound on the
amplitude\footnote{If there are sub-dominant
$\overline{\nu}_{i}K^{+}$ modes with branching ratio $R$, the right
side of Eq. (\ref{e42}) should be divided by $\sqrt{1+R}$.}:  
\begin{eqnarray} 
\hat{A}(\overline{\nu}_{\ell})\,\leq\,1\times10^{-24}\,\mbox{GeV}^{-1} 
\label{e42}
\end{eqnarray} 
which holds for both SU(5) and SO(10).  Recent theoretical analyses based 
on LEP-limit on Higgs mass $(\gtrsim 114$ GeV), together with certain 
assumptions about MSSM parameters (as in CMSSM) and/or constaint from 
muon g-2 anomaly \cite{g2} suggest that 
$\tan\beta\gtrsim$ 3 to 5 \cite{Arnowitt}. In the interest of getting a 
conservative upper limit on proton lifetime, we will therefore use, as 
a conservative lower limit, $\tan\beta\geq 3$. We will however exhibit 
relevant results often as a function of $\tan\beta$ and exhibit proton 
lifetimes corresponding to higher values of $\tan\beta$ as well. 
For minimal SU(5), using
Eq.(\ref{e36}) and, conservatively $\tan\beta\geq 3$,  
one obtains a lower limit on $M_{HC}$  given by: 
\begin{eqnarray} 
M_{HC}\,\geq\,5.5\times10^{16}\,\mbox{GeV}\,\,\,(\mbox{SU}(5)) 
\label{e43}
\end{eqnarray}
At the same time, higher values of $M_{HC}>3\times10^{16}$ GeV do  not
go very well with gauge coupling unification. Thus we already see a 
conflict, in the case of minimal SUSY SU(5), between the experimental 
limit on proton lifetime on the one hand, and coupling unification 
and constraint on $\tan\beta$ on the other hand. To see this conflict 
another way, if we keep $M_{HC}\leq3\times10^{16}$ GeV (for the sake of 
coupling unification) we  obtain from Eq.(\ref{e36}):
$\hat{A}(\mbox{SU}(5))\geq1.9\times10^{-24}\,\mbox{GeV}^{-1}(\tan\beta/3)$.
Using Eq. (\ref{e40}), this in turn implies that 
\begin{eqnarray} 
\Gamma^{-1}(p\,\rightarrow\,\overline{\nu}K^+)\,\leq\,
   0.6\times10^{33}\,\mbox{yrs} 
   \times(3/\tan\beta)^2\,\,\,\,\,(\mbox{SU}(5))
\label{e44} 
\end{eqnarray} 
For $\tan\beta \geq 3$, a lifetime of $0.7\times 10^{33}$ years is thus  
a conservative upper limit. In practice, it is unlikely  that
all the uncertainties, including these in $M_{HC}$ and $\tan\beta$, 
would stretch in
the same direction to nearly extreme values so as to prolong proton
lifetime.  A more reasonable upper limit, for minimal SU(5), thus
seems to  be:
$\Gamma^{-1}(p\rightarrow\overline{\nu}K^+)(SU(5))\leq(0.3)\times10^{33}$
yrs.  Given the experimental lower limit (Eq.(\ref{e41})), we  see
that minimal SUSY SU(5) is already excluded (or strongly disfavored) 
by  proton decay-searches.  We have of course noted in 
Sec.\ref{Mass} that SUSY  SU(5) does not go well with  neutrino oscillations
observed at  SuperK. 
 
Now, to discuss proton decay in the context of supersymmetric SO(10), it 
is necessary to discuss first the mechanism for doublet-triplet 
splitting.  Details of 
this discussion may be found in Ref.\cite{BabuWilczekPati}.  A synopsis is
presented in the Appendix.

\subsection{Proton Decay in Supersymmetric SO(10)}

The calculation of the amplitudes $\hat{A}_{std}$ and $\hat{A}_{new}$
for the standard  and the new operators for the SO(10) model, are
given in detail in Ref.\cite{BabuWilczekPati}.  Here, I will present only the
results.  It is found that  the four amplitudes
$\hat{A}_{std}(\overline{\nu}_\tau K^{+})$,
$\hat{A}_{std}(\overline{\nu}_{\mu}K^{+})$,
$\hat{A}_{new}(\overline{\nu}_{\tau}K^{+})$ and
$\hat{A}_{new}(\overline{\nu}_{\mu}K^{+})$ are in fact very comparable
to each other, within about a factor of two to five, either way. Since there
is no reason to expect a near cancellation between the standard and
the new operators, especially for both $\overline{\nu}_{\tau}K^{+}$
and $\overline{\nu}_{\mu}K^{+}$ modes, we expect the net amplitude
(standard + new) to be in the range exhibited by either one.
Following Ref.\cite{BabuWilczekPati}, I therefore present the 
contributions from the standard and the new operators separately. 

One important consequence 
of the doublet-triplet splitting mechanism for SO(10) outlined briefly 
in the appendix and in more detail in Ref.\cite{BabuWilczekPati} is that the 
standard d=5 proton decay operators become inversely proportional to
$M_{eff}\equiv [\lambda<45_H>]^2/M_{10'}\sim M_{X}^2/M_{10'}$, rather than 
to $M_{H_c}$. Here, $M_{10'}$ represents the mass of $10_H'$, that enters 
into the D-T splitting mechanism through effective coupling 
$\lambda 10_H 45_H 10_H'$ in the superpotential (see Appendix, 
Eq.(\ref{a1})). As noted in Ref.\cite{BabuWilczekPati}, $M_{10'}$ can be 
naturally suppressed (due to flavor symmetries) compared to $M_X$, and 
thus $M_{eff}$ correspondingly larger than $M_X$ by 
even one to three orders of 
magnitude. It should be stressed that $M_{eff}$ does not represent the 
physical masses of the color triplets or of the other particles in the 
theory. It is simply a parameter of order $M_X^2/M_{10'}$. {\it Thus 
larger values of $M_{eff}$, close to or even exceeding the Plank scale, 
do not 
in any way imply large corrections from quantum gravity}. Now accompanying 
the suppression due to $M_{eff}$, the standard proton decay amplitudes 
for SO(10) possess an intrinsic enhancement as well, compared to those for 
SU(5), owing primarily due to differences in their Yukawa couplings for the 
up sector (see Appendix C of Ref.\cite{BabuWilczekPati}). As a result 
of this enhancement, combined with the suppression due to higher values 
of $M_{eff}$, a typical standard $d=5$ amplitude for SO(10) is given by 
(see Appendix C of Ref.\cite{BabuWilczekPati})
\begin{displaymath} 
\hat{A}(\bar{\nu}_\mu K^+)_{std}^{SO(10)}\approx(h_{33}^2/M_{eff})
(2\times10^{-5}), 
\end{displaymath}
which should be compared with 
$\hat{A}(\bar{\nu}_\mu K^+)_{std}^{SU(5)}\approx(1.9\times10^{-8})
(\tan\beta/M_{H_c})$ (see Eq.(\ref{e36})). Note, taking $h_{33}^2\approx 1/4$, 
the ratio of a typical SO(10) over SU(5) amplitude is given by 
$(M_{H_c}/M_{eff})(88)(3/\tan\beta)$. Thus the enhancement by a factor of 
about 88 (for $\tan\beta=3$), of the SO(10) compared to the 
SU(5) amplitude, is compensated in 
part by the suppression that arises from $M_{eff}$ being larger than 
$M_{H_c}$. 

In addition, note that in contrast to the case of SU(5), the SO(10) 
amplitude does not depend {\it explicitly} on $\tan\beta$. The reason is 
this: if the fermions acquire masses only through the $10_H$ in SO(10), 
as is well known, the up and down quark Yukawa couplings will be equal. 
By itself, it would lead to a large value of $\tan\beta=m_t/m_b\approx 60$ 
and thereby to a large enhancement in proton decay amplitude. Furthermore, 
it would also lead to the bad relations: $m_c/m_s=m_t/m_b$ and $V_{CKM}=1$. 
However, in the presence of additional Higgs multiplets, in particular with 
the mixing of $(16_H)_d$ with $10_H$ (see Appendix and 
Sec.\ref{Understanding}), (a) $\tan\beta$ can get lowered to values like 
3-20, (b) fermion masses get contributions from both $<16_H>_d$ and 
$<10_H>$, which correct all the bad relations stated above, and 
simultaneously (c) the explicit dependence of $\hat{A}$ on  
$\tan\beta$ disappears. It reappears, however, through restriction on 
threshold corrections, discussed below.

Although $M_{eff}$ can far exceed $M_X$, it still gets bounded from above 
by demanding that coupling unification, as observed 
\footnote{
  For instance, in the absence of GUT-scale threshold corrections, the 
  MSSM value of $\alpha_3(m_Z)_{MSSM}$, assuming coupling unification,  
  is given by  $\alpha_3(m_Z)_{MSSM}^\circ=0.125-0.13$ \cite{Langacker}, 
  which is about 5-8\% higher than the observed value: 
  $\alpha_3(m_Z)_{MSSM}^\circ=0.118-0.003$ \cite{ParticleDataGroup}. 
  We demand that this discrepancy should be accounted for accurately 
  by a net {\it negative} contribution from D-T splitting and from 
  ``other'' threshold corrections (see Appendix, Eq.(\ref{a4})), without 
  involving large cancellations. That in fact does happen for the 
  minimal Higgs system $(45, 16, \overline{16})$ 
  [see Ref.\cite{BabuWilczekPati}].    
},
should emerge as a natural prediction of the theory as opposed to being 
fortuitous. That in turn requires that there be no large (unpredicted) 
cancellation between GUT-scale threshold corrections to the gauge 
couplings that arise from splittings within different multiplets as well 
as from Plank scale physics. Following this point of view, we have 
argued (see Appendix) that the net ``other'' threshold corrections 
to  $\alpha_3(m_Z)$ arising from the Higgs (in our case $45_H$, $16_H$ and 
$\overline{16}_H$) and the gauge multiplets should be negative, but 
conservatively and quite plausibly no more than about 10\%. This in 
turn restricts how big can be the threshold corrections to 
$\alpha_3(m_Z)$ that arise from (D-T) splitting (which is positive). 
Since the latter is proportional to $\ln(M_{eff}\cos\gamma/M_{X})$ 
(see Appendix), we thus obtain an upper limit on $M_{eff}\cos\gamma$. 
For the simplest model of D-T splitting presented in 
Ref.\cite{BabuWilczekPati} and in the Appendix (Eq.(\ref{a1})), one 
obtains: $\cos\gamma\approx(\tan\beta)/(m_t/m_b)$. An upper limit on 
$M_{eff}\cos\gamma$ thus provides an upper limit on $M_{eff}$ which is 
inversely proportional to $\tan{\beta}$. In short, our demand of 
natural coupling unification, together with the simplest model of 
D-T splitting, introduces an implicit dependence on $\tan\beta$ into 
the lower limit of the SO(10) - amplitude - i.e. 
$\hat{A}(SO(10))\propto 1/M_{eff} \geq$ (a quantity) $\propto \tan\beta$. 
These considerations are reflected in the results given below.

Assuming $\tan{\beta}\ge 3$ and accurate coupling unification 
(as described above), 
one obtains for the case of MSSM, a conservative upper limit on 
$M_{eff} \le 2.7\times 10^{18}$ GeV $(3/\tan\beta)$ 
(see Appendix and Ref.\cite{BabuWilczekPati}). 
Using this upper limit, 
we obtain a lower limit for the standard proton decay amplitude given by
\begin{eqnarray} 
\hat{A}(\overline{\nu}_{\tau}K^{+})_{std}\,\geq\,\left[\begin{array}{c}
  {(7.8\times10^{-24}\,\mbox{GeV}^{-1})\,(1/6\,\mbox{-\,1/4)}
	\,\,\,\,\,\,\,\,\mbox{caseI}}\\
{(3.3\times10^{-24}\,\mbox{GeV}^{-1})\,(1/6\,\mbox{-\,1/2)}\,\,\,\,\,\,\,\,\mbox{case II}}\end{array}\right]
  \left(\begin{array}{c}\mbox{SO(10)/MSSM, with}\\
	 \tan\beta \geq 3 \end{array}\right).
\label{e45} 
\end{eqnarray} 
Substituting into Eq.(\ref{e40}) and adding the contribution from the
second  competing mode $\overline{\nu}_{\mu}K^{+}$, with a typical
branching ratio $R\approx0.3$, we obtain 
\begin{eqnarray} 
\Gamma^{-1}(\overline{\nu}K^{+})_{std}\,\leq\,\left[\begin{array}{c}
{(0.7\times10^{31}\,\mbox{yrs.})\,(1.6\,\mbox{-}\,0.7)}\\
{(1.5\times10^{31}\,\mbox{yrs.})\,(4\,\mbox{-}\,0.44)}\end{array}\right]\,\{64\,\mbox{-}\,1/64\}
  \left(\begin{array}{c}\mbox{SO(10)/MSSM, with}\\
	 \tan\beta \geq 3 \end{array}\right).
\label{e46}
\end{eqnarray} 
The upper and lower entries in Eqs.(\ref{e45}) and (\ref{e46}) 
correspond to the cases I and II of the fermion mass-matrix 
with the {\it extreme values} of $\epsilon'$
- i.e.  $\epsilon'=2\times10^{-4}$ and $\epsilon'=0$ - respectively,
(see Eq.(\ref{e33})).  The  uncertainty shown inside the square
brackets correspond to that in the  relative phases of the different
contributions.  The uncertainty of \{64 to 1/32\} arises from that in
$\beta_{H}$,  $(m_{\tilde{W}}/m_{\tilde{q}})$ and $m_{\tilde{q}}$ (see
Eq.(\ref{e39})). Thus we find that for MSSM embedded in SO(10), 
for the two extreme values of $\epsilon'$ (cases I and II) as mentioned  
above, the inverse partial proton decay rate  should satisfy:
\begin{eqnarray} 
\Gamma^{-1}(p\,\rightarrow\overline{\nu}K^{+})_{std}\,\leq\,
\left[\begin{array}{c}{0.7\times10^{31_{-1.7}^{+2.0}}\,\mbox{yrs.}}\\
{1.3\times10^{31^{+2.4}_{-1.86}}\,\mbox yrs.}\end{array}\right]\,
\leq\,\left[\begin{array}{c}{0.7\times10^{33}\,\mbox{yrs.}}\\
{3.7\times10^{33}\,\mbox{yrs.}}\end{array}\right]\,\,\,\,
 \left(\begin{array}{c}\mbox{SO(10)/MSSM, with}\\
	\tan\beta \geq 3 \end{array}\right)\,.
\label{e47} 
\end{eqnarray}
The central value of the upper limit in Eq.(\ref{e47}) corresponds to
taking the upper limit on $M_{eff}\leq 2.7\times10^{18}$ GeV, which is 
obtained by restricting threshold corrections as described above 
(and in the Appendix) and by setting (conservatively) 
$\tan\beta\geq 3$. The uncertainties of matrix element, 
spectrum and choice of phases are reflected in the
exponents.The uncertainty in the most sensitive entry of the 
fermion mass matrix -
i.e. $\epsilon'$ - is fully incorporated (as regards obtaining an upper
limit on the lifetime) by going from case I 
(with $\epsilon'=2\times 10^{-4}$)
to case II ($\epsilon'=0$). Note that this increases the lifetime by
almost a factor of six. Any non-vanishing intermediate value of $\epsilon'$
would only shorten the lifetime compared to case II. In this sense,
the larger of the two upper limits quoted above is rather conservative.
We see that the predicted upper limit for case I of MSSM (with the 
extreme value of $\epsilon'=2\times 10^{-4}$) is already 
in conflict with the empirical lower limit (Eq.(\ref{e42})) while 
that for case II i.e. $\epsilon'=0$ (with all the uncertainties 
stretched as mentioned above) is only about two times higher  
than the empirical limit.  

Thus the case of MSSM embedded in SO(10) is already tightly 
constrained, to the point of being rather disfavored, 
by the limit on proton lifetime in that all the parameters 
need to lie near their ``extreme'' ends so that it may be compatible 
with the empirical limit (see also results for other choices of 
parameters listed in Table 1). The constraint is of course augmented 
especially by our requirement of natural coupling unification which 
prohibits accidental large cancellation between different threshold 
corrections (see Appendix); and it will be even more severe, especially 
within the simplest mechanism of D-T splitting (as discussed in the 
Appendix), if $\tan\beta$ turns out to be larger than 5 (say). On the 
positive side, improvement in the current limit by a factor of even 
2 to 3 ought to reveal proton decay, otherwise the case of MSSM embedded 
in SO(10), would be clearly excluded. 

\subsection{The case of ESSM}

Before discussing the contribution of the new $d=5$ operators to proton 
decay, an interesting possibility, mentioned in the introduction, that 
would be especially relevant in the context of proton decay, if 
$\tan\beta$ is large, is worth noting. This is the case of the extended 
supersymmetric standard model (ESSM), which introduces an extra pair of 
vector-like families ($16+\overline{16}$ of SO(10)), at the TeV scale 
\cite{BabuJi,BabuPatiStrem}. Adding such complete SO(10)-multiplets 
would of course preserve coupling unification. 
From the point of view of adding extra families, 
ESSM seems to be the minimal and also the maximal extension of the 
MSSM, that is allowed in that it is compatible with (a) neutrino-counting, 
(b) precision electroweak tests, as well as (c) a semi-perturbative as 
opposed to non-perturbative gauge coupling unification 
\cite{BabuJi,BabuPatiStrem}. 
\footnote{
   For instance, 
   addition of two pairs of vector-like families at the TeV-scale, 
   to the three chiral families, would cause gauge couplings to become 
   non-perturbative below the unification scale.
} 
{\it The existence of two extra vector-like families can of course be tested 
at the LHC.} 

Theoretical motivations for the case of ESSM arise because, (a) it 
raises $\alpha_{\mbox{\scriptsize{unif}}}$ to a semi-perturbative value 
of 0.25 to 0.3, and therefore has a better chance to achieve 
dilaton-stabilization than the case of MSSM, for which 
$\alpha_{\mbox{\scriptsize{unif}}}$ is rather weak (only 0.04); 
and (b) owing to increased two-loop effects \cite{BabuJi,KoldaRussell}, 
it raises the unification 
scale $M_X$ to $(1/2-2)\times 10^{17}$GeV and thereby considerably reduces 
the problem of a mismatch \cite{DienesJCP} between the MSSM and the 
string unification scales (see Sec.\ref{Need}). A third feature relevant to 
proton decay is the following. In the absence of unification-scale 
threshold and Planck-scale effects, the ESSM value of $\alpha_3(m_Z)$ 
obtained by assuming gauge coupling unification, which we denote by 
$\alpha_3(m_Z)_{\mbox{\scriptsize{ESSM}}}^\circ$
is lowered to about $0.112-0.118$ \cite{BabuJi}, compared to 
$\alpha_3(m_Z)_{\mbox{\scriptsize{MSSM}}}^\circ \approx 0.125 - 0.13$. 

As explained in the appendix, the net result of these two effects - i.e. 
a raising of $M_X$ and a lowering of 
$\alpha_3(m_Z)_{\mbox{\scriptsize{ESSM}}}^\circ$ - is that for ESSM 
embedded in SO(10), $\tan\beta$ can span a wide range from 3 to even 30, 
and simultaneously the value or
the upper limit on $M_{eff}$ can range from 
$(60$ to $6)\times 10^{18}$GeV, in full accord with our criterion for 
accurate coupling unification discussed above. 

Thus, in contrast to MSSM, ESSM allows for larger values of $\tan\beta$ 
(like 20 or 30), without needing large threshold corrections, and 
simultaneously without conflicting with the limit on proton lifetime. 

To be specific, consider first the case of a moderately large $\tan\beta=20$ 
(say), for which one obtains $M_{eff}\approx 9\times 10^{18}$ GeV, with 
the ``other'' threshold correction $-\delta_3'$ being about 5\% 
(see Appendix for definition). In this case, one obtains: 
\begin{eqnarray} 
\Gamma^{-1}(\overline{\nu}K^{+})_{std}\,\approx\,
\left[\begin{array}{c}{(1.6-0.7)}\\
{(10-1)}\end{array}\right]\,\{64-1/64\}\, (7\times 10^{31}\,\mbox{yrs})
 \left(\begin{array}{c}\mbox{SO(10)/ESSM, with} 
        \\ \tan\beta=20\end{array}\right).
\label{e49-2} 
\end{eqnarray}
As before, the upper and lower entries correspond to cases I 
($\epsilon'=2\times 10^{-4}$) and II ($\epsilon'=0$) of 
the fermion mass-matrix (see Eq.(\ref{e33})). The uncertainty in the 
upper and lower entries in the square bracket of Eq.(\ref{e49-2})
corresponds to that in the relative phases of the different 
contributions for the cases I and II respectively, 
while the factor \{64-1/64\} corresponds to uncertainties 
in the SUSY spectrum and the matrix element (see Eq.{\ref{e39}}). 

We see that by allowing for an uncertainty of a factor of $(30-100)$ 
jointly from the two brackets for Case I (and $(13-44)$ for Case II), 
proton lifetime arising from the standard operators 
would be expected to lie in the range of 
$(2.2-7.5)\times10^{33}$ yrs, for the case of ESSM embedded in SO(10),  
with $\tan\beta=20$. Such a range is compatible with present limits, 
but accessible to searches in the near future. 

The other most important feature of ESSM is that, by allowing for 
larger values of $M_{eff}$, especially for smaller values of 
$\tan\beta\approx 3$ to $10$ (say), {\it the contribution of the 
standard operators by itself can be perfectly consistent 
with present limit on proton lifetime even for almost central or 
``median'' values 
of the parameters pertaining to the SUSY spectrum, the relevant 
matrix element, $\epsilon'$ and the phase-dependent factor}. 

For instance, for ESSM, one obtains $M_{eff}\approx (4.5\times 
10^{19} \mbox{GeV}) (4/tan\beta)$, with the ``other'' threshold 
correction - $\delta_3'$ being about 5\% (see Appendix and 
Eq.(\ref{MeffEq})). Now, {\it combining} cases I 
($\epsilon'=2\times 10^{-4}$) and II ($\epsilon'=0$), we see that 
the square bracket in Eq.(\ref{e49-2}) which we will denote by 
[S], varies from 0.7 to 10, depending upon the relative phases 
of the different contributions and the values of $\epsilon'$. Thus 
as a ``median'' value, we will take $[S]_{med} \approx$ 2 to 6. 
The curly bracket \{64-1/64\}, to be denoted by \{C\}, represents 
the uncertainty in the SUSY spectrum and the matrix element 
(see Eq.(\ref{e39})). Again as a ``nearly central'' or ``median'' 
value, we will take $\{C\}_{med}\approx 1/6 \mbox{ to } 6$. Setting 
$M_{eff}$ as above we obtain
\begin{equation}\label{e50-2}
  \Gamma^{-1}(\bar{\nu}K^+)_{std}^{``median''} \approx 
	[S]_{med}\{C\}_{med} (1.8\times 10^{33} \mbox{ yrs}) 
	(4/\tan\beta)^2 (\mbox{SO(10)/ESSM}).
\end{equation}
Choosing a few sample values of the effective parameters [S] and 
\{C\}, with low values of $\tan\beta = 4$ to 10, the corresponding 
values of $\Gamma^{-1}(\bar{\nu}K^+)$, following from Eq.(\ref{e50-2}), 
are listed below in Table 1. 

Note that ignoring contributions from the new d=5 operators for a moment 
\footnote{
  As I will discuss in the next section, we of course expect the new 
  d=5 operators to be important and significantly infuence proton 
  lifetime (see e.g. Table 2). Entries in Table 1 could still represent 
  the actual expected values of proton lifetimes, however, if the parameter K 
  defined in \ref{Preliminaries} (also see \ref{Contribution}) happens 
  to be unexpectedly small ($\ll 1$). 
},
the entries in Table 1 represent {\it a very plausible range of values}
for the proton lifetime, for the case of ESSM embedded in SO(10), with 
$\tan\beta\approx 3 \mbox{ to } 10$ (say), {\it rather than upper limits 
for the same}. This is because they are obtained for ``nearly central'' 
or ``median'' values of 
\vspace*{12pt}
\begin{center}
 {\bf TABLE 1. PROTON LIFETIME, BASED ON CONTRIBUTIONS FROM ONLY THE 
  STANDARD OPERATORS FOR THE CASE OF ESSM EMBEDDED IN SO(10), WITH 
  PARAMETERS BEING IN THE ``MEDIAN'' RANGE}
\end{center}
\noindent
\begin{tabular}{|c|c|c|c|}
\hline
  $\tan\beta=4$	& $\tan\beta=4$	& $\tan\beta=10$ & $\tan\beta=10$ \\ 
  \hspace{0pt}[S]=2.7 & [S]=6	& [S]=5.4	& [S]=6	\\
  \{C\}=1/2 to 2& \{C\}=1/6 to 1& \{C\}=1 to 6	& \{C\}=1 to 4  \\
\hline
  $\Gamma^{-1}(\bar{\nu}K^+)_{ESSM}^{std}\approx$ & 
   	$\Gamma^{-1}(\bar{\nu}K^+)_{ESSM}^{std}\approx$ &
		$\Gamma^{-1}(\bar{\nu}K^+)_{ESSM}^{std}\approx$ &
		$\Gamma^{-1}(\bar{\nu}K^+)_{ESSM}^{std}\approx$   \\
 $(2.5\mbox{ to }10)\times 10^{33}$ yrs & 
	$(1.8\mbox{ to }11)\times 10^{33}$ yrs & 
	$(1.6\mbox{ to }10)\times 10^{33}$ yrs & 
	$(1.8\mbox{ to }7.3)\times 10^{33}$ yrs \\
\hline
\end{tabular}

\vspace*{12pt}
\noindent
the parameters represented by the values of 
[S]$\approx$ 2 to 6 and \{C\}$\approx$ 1/6 to 6, as discussed above. 
For instance, consider the cases \{C\}=1 and \{C\}=1/6 respectively, 
both of which (as may be inferred from the table) can quite plausibly 
yield proton lifetimes in the range of $10^{33}$ to $10^{34}$ yrs. Now 
\{C\}=1 corresponds, e.g., to $\beta_H=0.014$GeV$^3$ (the central value 
of Ref.\cite{Aoki}) $m_{\tilde{q}}=1.2$ TeV and 
$m_{\tilde{W}}/m_{\tilde{q}}=1/6$ (see Eq.(\ref{e39})), while that of 
\{C\}=1/6 would correspond, for example, to $\beta_H=0.014$GeV$^3$, 
with $m_{\tilde{q}}\approx 600$GeV and 
$m_{\tilde{W}}/m_{\tilde{q}}\approx 1/5$. {\it In short, for the case 
of ESSM, with low values of $\tan\beta\approx$ 3 to 10 (say), squark 
masses can be well below 1 TeV, without conflicting with present limit 
on proton lifetime}. This feature is not permissible within MSSM embedded 
in SO(10).  

Thus, confining for a moment to the 
standard operators only, if ESSM represents low-energy physics, 
and if $\tan\beta$ is rather small (3 to 10, say), we do not have to 
stretch at all the uncertainties in the SUSY spectrum and the matrix elements 
to their extreme values (in contrast to the case of MSSM)
in order to understand why proton decay has not 
been seen as yet, and still can be optimistic that it ought to be 
discovered in the near future, with a lifetime $\leq 10^{34}$ years.  
The results for a wider variation of the parameters are listed in 
Table 2, where contributions of the new d=5 operators are also shown.   

It should also be remarked that if in the unlikely event, all the 
parameters (i.e. $\beta_H$, $(m_{\tilde{W}}/m_{\tilde{q}})$, 
$m_{\tilde{q}}$ and the phase-dependent factor) happen to be closer 
to their extreme values so as to extend proton lifetime, and if 
$\tan\beta$ is small ($\approx 3$ to 10, say) and at the same time 
the value of $M_{eff}$ is close to its allowed upper limit (see 
Appendix), the standard d=5 operators by themselves would tend to 
yield proton lifetimes exceeding even (1/3 to 1)$\times10^{35}$ years for the 
case of ESSM, (see Eq.(\ref{e49-2}) and Table 2). 
In this case (with the parameters having nearly extreme 
values), however, as I will discuss shortly, the contribution of the 
new d=5 operators  related to neutrino masses (see Eq.(\ref{e35})), 
would dominate and quite naturally yield lifetimes bounded above in 
the range of $(1-10)\times 10^{33}$ years (see Sec.\ref{Contribution} 
and Table 2). {\it Thus in the presence of the new operators,  
the range of $(10^{33}-10^{34})$ years for proton lifetime is not 
only very plausible but it also provides a conservative upper limit, 
for the case of ESSM embedded in SO(10)}.

\subsection{Contribution from the new d=5 operators}\label{Contribution}

As mentioned in Sec.\ref{Preliminaries}, for supersymmetric G(224)/SO(10), 
there very likely exists a new set of d=5 operators, related to neutrino 
masses, which can induce proton decay (see, Eq.(\ref{e41})). The decay 
amplitude for these operators for the leading mode (which in this case 
is $\bar{\nu}_\mu K^+$) becomes proportional to the quantity 
$P\equiv \{(f_{33})_\nu\langle
	\overline{16}_H\rangle/M\} h_{33}K/(M_{16}\tan\gamma)$, 
where $(f_{33})_\nu$ and $h_{33}$ are the effective  couplings defined in  
Eqs.(\ref{e7}) and (\ref{e15}) respectively, and $M_{16}$ and $\tan\gamma$ 
are defined in the Appendix. The factor K, defined by 
$(f_{33})_p\equiv (f_{33})_\nu K$, is expected to be of order unity 
(see Sec.\ref{Expectations}A for the origin of K). As a plausible 
range, we take $K\approx 1/3$ to 2. 
Using $M_{16}\tan\gamma=\lambda'\langle\overline{16}_H\rangle$ 
(see Appendix), and  $h_{33}\approx 1/2$ (given by top mass), one gets: 
$P\approx ((f_{33})_\nu/M)(1/2\lambda')K$. Here M denotes the string or the 
Planck scale (see Sec.\ref{Mass} and footnote 2); thus 
$M\approx (1/2 -1)\times 10^{18}$GeV; and $\lambda'$ is a quartic coupling 
defined in the appendix. Validity of perturbative calculation suggests that 
$\lambda'$ should not much exceed unity, while other considerations suggest 
that $\lambda'$ should not be much less than unity  either 
(see Ref.\cite{BabuWilczekPati}, Sec.6E). Thus, a plausible range for 
$\lambda'$ is given by $\lambda'\approx(1/2 -\sqrt{2})$. 
(Note it is only the upper limit on $\lambda'$ 
that is relevant to obtaining an upper limit on proton lifetime). Finally, 
from consideration of $\nu_\tau$ mass, we have $(f_{33})_\nu\approx 1$ 
(see Sec.\ref{Mass}). We thus obtain: 
$P\approx (5\times10^{-19}\mbox{GeV}^{-1})(1/\sqrt{2}$ to 4)K. 
Incorporating a 
further uncertainty by a factor of (1/2 to 2) that arises due to choice 
of the relative phases of the different contributions 
(see Ref.\cite{BabuWilczekPati}), the effective 
amplitude for the new operator is given by  
\begin{equation}\label{e51}
   \hat{A}(\bar{\nu}_\mu K^+)_{new}\approx (1.5\times 10^{-24}\mbox{GeV}^{-1}) 
		(1/2\sqrt{2}\,\, \mbox{to}\,\, 8)K
\end{equation}
Note that this new contribution is independent of $M_{eff}$; {\it thus it is 
the same for ESSM as it is for MSSM, and it is independent of $\tan\beta$}. 
Furthermore, it turns out that the new contribution is also insensitive 
to $\epsilon'$; thus it is nearly the same for cases I and II of the 
fermion mass-matrix. Comparing Eq.(\ref{e51}) with Eq.(\ref{e45}) we see 
that the new and the standard operators 
are typically quite comparable to one another. Since there is no reason 
to expect near cancellation between them (especially for both 
$\bar{\nu}_\mu K^+$ and $\bar{\nu}_\tau K^+$ modes), we expect the net 
amplitude (standard+new) to be in the range exhibited by either one. 
It is thus useful to obtain the inverse decay rate assuming as if the new 
operator dominates. Substituting Eq.(\ref{e51}) into Eq.(\ref{e40}) 
and allowing for the presence of the $\bar{\nu}_\tau K^+$ mode with 
an estimated branching ratio of nearly 0.4 (see Ref.\cite{BabuWilczekPati}), 
one obtains
\begin{eqnarray}
\Gamma^{-1}(\overline{\nu}K^{+})_{new}\,\approx\,(1\times10^{31}\,
  \mbox{yrs})\,[8\,\mbox{-}\,1/64]\,\{64\,\mbox{-}\,1/64\}
  (K^{-2}\approx 9\,\,\mbox{to}\,\, 1/4)\,. 
\label{e52}
\end{eqnarray}
The square bracket represents the uncertainty reflected in Eq.(\ref{e51}), 
while the curly bracket corresponds to that in the SUSY spectrum and 
matrix element (Eq.(\ref{e39})). 
Allowing for a net uncertainty at the upper end by as
much as a factor of  100
to 600 (say), arising jointly from the {\it three brackets} 
in Eq.(\ref{e52}), which can be realized by keeping the SUSY-spectrum 
and the matrix element in the ``nearly-central'' or ``intermediate'' 
range (see below), 
the new operators related  to neutrino masses, by  themselves, 
lead to a proton decay lifetime given by: 
\begin{equation} 
\Gamma^{-1}(\overline{\nu}K^{+})^{Median}_{new}\,
	\approx\,(0.7\,\mbox{-}\,5)\times10^{33}\,\mbox{yrs.}\,\,\mbox{(SO(10)
or string G(224))} (\mbox{ Indep. of }\tan\beta).
\label{e53} 
\end{equation}
  For instance, taking the curly bracket in Eq.(\ref{e52}) to be 
  $\approx 4$ to 10 (say) (corresponding for example, to 
  $\beta_H=0.012$ GeV$^3$, 
  $(m_{\tilde{W}}/m_{\tilde{q}})\approx 1/10$ to 1/12 and 
  $m_{\tilde{q}}\approx (1$ to 1.3)(1.2 TeV)), instead of its extreme 
  value of 64, and setting the square bracket in Eq.(\ref{e52}) 
  to be $\approx 6$, and 
  $K^{-2}\approx 9$, which are quite plausible, we obtain: 
  $\Gamma^{-1}(\bar{\nu}K^+)_{new}\approx (2.2-4)\times 10^{33}$ yrs; 
  independently of $\tan\beta$, for both MSSM and ESSM. 
  Proton lifetime for other choices 
  of parameters, which lead to similar conclusion, are listed in Table 2. 

It should be stressed that the standard $d=5$ operators 
(mediated by the color-triplets in the $10_H$ of SO(10)) may naturally be
absent for a string-derived G(224)-model (see e.g. Ref.\cite{Antoniadis} 
and \cite{FaraggiHalyo}), but the new $d=5$ operators,
related to the Majorana masses of the RH neutrinos and the CKM
mixings, should very likely be present for such a model, as much 
as for SO(10). These would induce proton decay 
\footnote{
  In addition, quantum gravity induced d=5 operators are also expected 
to be present at some level, depending upon the degree of suppression of 
these operators due to flavor symmetries (see e.g. Ref.\cite{JCPProton}).
}. 
{\it Thus our
expectations for the proton decay lifetime (as shown in Eq. (\ref{e53}))
and the prominence of the $\mu^{+}K^{0}$ mode (see below) hold for a
string-derived G(224)-model, just as they do for SO(10)}. For a 
string - G(224) - model, however, the new d=5 operators would be 
essentially the sole source of proton decay. 

Nearly the same situation emerges for the case of ESSM embedded in 
G(224) or SO(10), with low $\tan\beta (\approx 3$ to 10, say), 
especially if the parameters (including $\beta_H$, 
$m_{\tilde{W}}/m_{\tilde{q}}$, $m_{\tilde{q}}$, the phase-dependent 
factor as well as $M_{eff}$) happen to be somewhat closer to their 
extreme values so as to extend proton lifetime. In this case, as 
noted in the previous sub-section, the contribution of the 
standard d=5 operators would be suppressed; and proton decay would 
proceed primarily via the new operators with a lifetime quite 
naturally in the range of $10^{33}-10^{34}$ years, as exhibited above. 

\subsection{The Charged Lepton Decay Modes $(p\rightarrow\mu^{+}K^{0}$  
	and $p\rightarrow e^+\pi^0$)}
\label{TheCharged}

I now note a distinguishing feature of the SO(10) or the
G(224) model presented here. 
Allowing for  uncertainties in the way the standard and the
new operators can combine  with each other for the three leading modes
i.e. $\overline{\nu}_{\tau}K^{+}$,  $\overline{\nu}_{\mu}K^{+}$ and
$\mu^{+}K^{0}$, we obtain (see Ref. \cite{BabuWilczekPati} for  details): 
\begin{eqnarray} 
B(\mu^{+}K^{0})_{std+new}\,\approx\,\left[1\%\,\,
\mbox{to}\,\,50\%\right]\,\kappa\,\,\,\,\mbox{(SO(10)
or string G(224))}
\label{e54} 
\end{eqnarray} 
where $\kappa$ denotes the ratio of the squares of relevant matrix
elements  for the $\mu^{+}K^{0}$ and $\overline{\nu}K^{+}$ modes. In
the absence of a reliable lattice calculation for the $\bar{\nu}K^{+}$
mode, one should remain open to the possibility  of
$\kappa\approx1/2$ to 1 (say). 
 We find that for a
large range of parameters, the branching ratio $B(\mu^{+}K^{0})$ can
lie in the range of 20 to 40\% (if $\kappa \approx1$). This prominence of
the $\mu^{+}K^{0}$ mode for the SO(10)/G(224) model
 is primarily due to contributions from the new d=5 
operators. This contrasts sharply with the minimal SU(5) model, in
which the $\mu^{+}K^{0}$ mode is expected to have a branching ratio of
only about $10^{-3}$. In short, prominence of the $\mu^{+}K^{0}$ mode,
if seen, would clearly show the relevance of the new operators, and
thereby reveal the proposed link between neutrino masses and proton
decay \cite{BPW1}.

While the d=5 operators as described here (standard and new) would lead to 
highly suppressed $e^+\pi^0$ mode, for MSSM or ESSM embedded in SO(10), 
the gauge-mediated d=6 operators, can still give proton decay into 
$e^+\pi^0$ with an inverse rate $\approx 10^{35.3\pm 1.5}$ years, 
which can be as short as about $10^{34}$ yrs. Thus, even within 
supersymmetric unification, the $e^+\pi^{0}$ mode may well be a 
prominent one, competing favorably with (even) the $\bar{\nu}K^{+}$ 
mode.

\subsection{Section Summary}

In summary, our study of proton decay has been carried out within the
supersymmetric 
SO(10) or the G(224)-framework\footnote{As described in 
Secs.\ref{Need}, \ref{Mass} and \ref{Understanding}.}, 
with special attention paid to its dependence on fermion masses
and threshold effects. A representative set of results corresponding 
to different choices of parameters is presented in Tables 1 and 2. 
The study strongly suggests that, for either MSSM or ESSM embedded in 
SO(10) or G(224), an upperlimit on proton lifetime is given by
\begin{eqnarray}
\tau_{proton}\,\leq\,(1/2\,\mbox{-}\,1)\times10^{34}\,\,\mbox{yrs}\,,
\label{e51-2}
\end{eqnarray}
with $\overline{\nu}K^{+}$ being the dominant decay mode, and 
$\mu^+K^{0}$ being prominent. Although
there are uncertainties in the matrix element, in the SUSY-spectrum, 
in the phase-dependent factor, $\tan\beta$ 
and in certain sensitive elements of the fermion mass matrix,
notably $\epsilon'$ (see Eq.(\ref{e47}) for predictions in cases I
versus II), this upper limit is obtained, for the case of MSSM 
embedded in SO(10), by allowing for a generous
range in these parameters and stretching all of them in the same
direction so as to extend proton lifetime. In this sense, while the
predicted lifetime spans a wide range, the upper limit quoted above,  
in fact more like $3\times 10^{33}$ yrs, is most conservative, 
for the case of MSSM (see Eq.(\ref{e47}) and Table 1).
It is thus tightly constrained already by the empirical lower limit 
on $\Gamma^{-1}(\overline{\nu}K^+)$ of $1.6\times 10^{33}$ yrs. 
For the case of ESSM embedded in SO(10), the standard d=5 operators 
are suppressed compared to the case of MSSM; as a result, by themselves 
they
can naturally lead to lifetimes in the range of $(3-10)\times 10^{33}$ 
yrs., for nearly central values of the parameters pertaining to the 
SUSY-spectrum and the matrix element (see Eq.(\ref{e50-2})) and Table 1. 
Including the contribution of the new d=5 operators, and allowing 
for a wide variation of the parameters mentioned above, one finds that 
the range of $(10^{33}-10^{34})$ yrs for proton lifetime is not only 
very plausible but it also provides a rather conservative upper limit, 
for the case of ESSM embedded in either SO(10) or G(224) 
(see Sec.\ref{Contribution} and Table 2). Thus our study provides 
a clear reason to expect 
that the discovery of proton decay should be imminent for the case 
of ESSM, and even more so for that of MSSM.  The implication of
this prediction for a next-generation detector is emphasized in the
next section. 

\section{Concluding Remarks}
\label{Conclusions}

The preceding sections show that, but for one missing piece 
-- proton decay -- the evidence in support of grand unification 
is now strong. It includes: (i) the
observed family-structure, (ii) the meeting of the gauge couplings,
(iii) neutrino-oscillations, (iv) the intricate pattern of the masses
and mixings of all fermions, including the neutrinos, and
 (v) the need for $B-L$ as a generator, to implement
baryogenesis. Taken together, these not only favor grand unification
but in fact select out a particular route to such unification, based
on the ideas of supersymmetry, SU(4)-color and left-right
symmetry. Thus they point to the relevance of an effective
string-unified G(224) or SO(10)-symmetry.

Based on a systematic study of proton decay within the supersymmetric 
SO(10)/G(224)-framework \cite{BabuWilczekPati}, which is clearly 
favored by the data, and an update as presented here, I have argued  
that a conservative upper limit on the proton
lifetime is about (1/2 - 1)$\times10^{34}$ yrs. for the case of either 
MSSM or ESSM, embedded in SO(10) or a string - G(224). 

So, unless the fitting of all the pieces listed 
above is a mere coincidence, and I
believe that that is highly unlikely, discovery of proton decay should
be around the corner. In particular, as mentioned in the Introduction,
we expect that candidate events should very likely be observed 
in the near future
already at SuperK. However, allowing for the possibility that proton
lifetime may well be near the upper limit or value stated above, a
next-generation detector providing a net gain in sensitivity by a
factor five to ten, compared to SuperK, would be needed to produce real
events and distinguish them unambiguously from the background. Such an
improved detector would of course be essential to study the branching
ratios of certain crucial though (possibly) sub-dominant decay modes 
such as the $\mu^{+}K^{0}$ and $e^+\pi^0$ as mentioned in 
Sec.\ref{TheCharged}.

The reason for pleading for such improved searches is that proton
decay would provide us with a wealth of knowledge about physics at
truly short distances ($<10^{-30}$ cm), which cannot be gained by any
other means. Specifically, the observation of proton decay, at a rate
suggested above, with $\overline{\nu}K^{+}$ mode being dominant, would
not only reveal the underlying unity of quarks and leptons  but also
the relevance of supersymmetry. It would also confirm a unification of
the fundamental forces at a scale of order $2\times10^{16}$ GeV.
Furthermore, prominence of the $\mu^{+}K^{0}$ mode, if seen, would
have even deeper significance, in that in addition to supporting  the
three features mentioned above, it would also reveal the  link between
neutrino masses and proton decay, as discussed in Sec.\ref{Expectations}. 
{\it In
this sense, the role of  proton decay in probing into physics at the
most fundamental level is unique }. In view of  how valuable
such a probe would be and the fact that the predicted  upper limit on
the  proton lifetime is at most a factor of three to six higher than the
empirical lower limit, the argument in favor of building an improved
detector seems compelling.

To conclude, the discovery of proton decay would undoubtedly
constitute a landmark in the history of physics. It would provide the
last, missing piece of gauge unification and would shed light on how such
a unification may be extended to include gravity in the context of a 
deeper theory.

\vskip1.5em

{\bf Acknowledgments :} I would like to thank Kaladi S. Babu and
Frank Wilczek for a most enjoyable collaboration, and Joseph Sucher
for valuable discussions. Discussions with Kaladi S. Babu in updating 
the results of the previous study have been most helpful. I would 
like to thank Antonio Zichichi and Gerard 't Hooft for arranging 
a stimulating school at Erice and also for the kind hospitality. 
The  research presented here is supported in part by DOE
grant no. DE-FG02-96ER-41015.

\appendix 
 
\section*{A Natural Doublet-Triplet Splitting Mechanism in
SO(10)} 
 
\setcounter{equation}{0} 
 
\renewcommand{\theequation}{A\arabic{equation}}

In supersymmetric SO(10), a  natural doublet--triplet splitting can be
achieved by coupling the adjoint  Higgs ${\bf 45_{H}}$ to a ${\bf
10_{H}}$ and a ${\bf 10'_{H}}$, with  ${\bf 45_{H}}$ acquiring a
unification--scale VEV in the $B$-$L$ direction  \cite{DimWilczek}:
$\langle{\bf 45_{H}}\rangle=(a,a,a,0,0)\times\tau_{2}$ with
$a\sim{M}_{U}$.  As discussed in Section \ref{Understanding}, 
to generate  CKM mixing
for fermions we require  $({\bf 16_{H}})_d$ to acquire a VEV of the
 electroweak scale. To ensure accurate gauge coupling unification,
the effective low energy theory should not contain split multiplets
beyond those of MSSM. Thus the MSSM Higgs doublets must  be linear
combinations of the SU(2$)_{L}$ doublets in ${\bf 10_{H}}$  and ${\bf
16_{H}}$. A simple set of superpotential terms that ensures this  and
incorporates doublet-triplet splitting is \cite{BabuWilczekPati}: 
\begin{eqnarray} 
W_{H}\,=\,\lambda\,{\bf 10_{H}\,45_{H}\,10'_{H}}\,+\,M_{10}\,{\bf
10'_{H}}^{2}\,+\,\lambda'\,\overline{\bf 16}_{H}\,\overline{\bf
16}_{H}\,{\bf 10}_{H}\,+\,M_{16}\,{\bf 16}_{H}\overline{\bf 16}_{H}\,.
\label{a1} 
\end{eqnarray}
A complete superpotential for ${\bf 45_{H}}$, ${\bf 16_{H}}$,  ${\bf
\overline{16}_{H}}$, ${\bf 10}_{H}$, ${\bf 10}'_{H}$ and possibly
other fields,  which  ensure that (a) ${\bf 45_{H}}$, ${\bf 16_{H}}$  and
${\bf \overline{16}_{H}}$ acquire unification  scale VEVs with
$\langle{\bf 45_{H}}\rangle$ being  along the $(B$-$L)$ direction; 
(b) 
that exactly two Higgs doublets  $(H_{u},H_{d})$ remain light, with
$H_{d}$ being a linear combination  of $({\bf 10_{H}})_{d}$ and $({\bf
16_{H}})_{d}$; 
and (c) there are  no unwanted pseudoGoldstone bosons,
can be constructed. With $\langle{\bf 45_{H}}\rangle$ in the $B$-$L$
direction, it  does not contribute to the Higgs doublet mass matrix, so
one pair of  Higgs doublet remains light, while all triplets acquire
unification  scale masses.  The light MSSM Higgs doublets are 
\cite{BabuWilczekPati}
\begin{eqnarray} 
H_{u}\,=\,{\bf 10}_{u}\,,\,\,\,\,H_{d}\,=\,\cos\gamma\,{\bf
10}_{d}\,+\,\sin\gamma\,{\bf 16}_{d}\,,
\label{a2} 
\end{eqnarray}
with
$\tan\gamma\equiv\lambda'\langle{\bf\overline{16}_{H}}\rangle/M_{16}$.
Consequently, $\langle{\bf10}\rangle_{d}=(\cos\gamma)\,v_{d}$,
$\langle{\bf16}_{d}\rangle=(\sin\gamma)\,v_{d}$, with $\langle
H_{d}\rangle=v_{d}$ and $\langle{\bf 16}_{d}\rangle$ and $\langle{\bf
10}_{d}\rangle$ denoting the electroweak VEVs of those
multiplets. Note that $H_{u}$ is purely in ${\bf 10_H}$ and that
$\left \langle {\bf 10}_d \right \rangle^2 + \left \langle {\bf 16}_d
\right \rangle^2 = v_d^2$. This mechanism of doublet-triplet (DT)
splitting is the simplest for the minimal Higgs systems. 
It has the advantage that 
meets the requirements of both D-T splitting and CKM-mixing.  In turn,
it has three special consequences:

(i) It modifies the familiar SO(10)-relation
$\tan \beta \equiv v_u/v_d = m_t/m_b \approx 60$ to 
\footnote{
  It is worth noting that the simple relationship between $\cos\gamma$ 
  and $\tan\beta$ - i.e. $\cos\gamma\approx\tan\beta/(m_t/m_b)$ - 
  would be modified if the superpotential contains an additional term 
  like $\lambda''16_H\cdot 16_H\cdot 10_H'$, which would induce a 
  mixing between the doublets in $10'_d$, $16_d$ and $10_d$. That in 
  turn will mean that the upper limit on $M_{eff}\cos\gamma$ 
  following from considerations of threshold corrections (see below) 
  will not be strictly proportional to $\tan\beta$. I thank 
  Kaladi Babu for making this observation.
}:
\begin{eqnarray}
\tan \beta/\cos \gamma \approx m_t/m_b \approx 60
\end{eqnarray}
As a result, even
low to moderate values of $\tan \beta
\approx 3$ to 10 (say) are perfectly allowed in SO(10) (corresponding to
$\cos \gamma \approx 1/20$ to $1/6$).

(ii) The most important consequence of the DT-splitting mechanism
outlined above is this: In contrast to SU(5), for which the strengths
of the standard d=5 operators are proportional to $(M_{H_c})^{-1}$
(where $M_{H_C}\sim few \times 10^{16}$ GeV (see Eq. (\ref{e43})), for
the SO(10)-model, they become proportional to $M_{eff}^{-1}$, where
$M_{eff} =(\lambda a)^2/M_{10'} \sim M_X^2/M_{10'}$. As noted in 
Ref.\cite{BabuWilczekPati}, $M_{10'}$ can be
naturally smaller (due to flavor symmetries) than $M_X$ and thus
$M_{eff}$ correspondingly larger than $M_{X}$ by even one to 
three orders of magnitude. Now the proton decay amplitudes for
SO(10) in fact possess an intrinsic enhancement compared to those for
SU(5), owing primarily due to differences in their Yukawa couplings
for the up sector (see Appendix C in Ref. \cite{BabuWilczekPati}). As a result,
these larger values of $M_{eff}\sim(10^{18}-10^{19})$ GeV are in fact needed
for the SO(10)-model to be compatible with the observed limit on the
proton lifetime. At the same time, being bounded above 
by considerations of threshold effects (see below),
they allow optimism as regards future observation of proton decay.

(iii) $M_{eff}$ gets bounded above by considerations of coupling
unification and GUT-scale threshold effects as follows. Let us 
recall that in the absence of unification-scale threshold and 
Planck-scale effects, the MSSM value of $\alpha_3(m_Z)$ in the 
$\overline{\mbox{MS}}$ scheme, obtained by assuming gauge coupling 
unification, is given by 
$\alpha_3(m_Z)_{\mbox{\scriptsize{MSSM}}}^\circ = 0.125 - 0.13$ 
\cite{Langacker}. 
This is about  5 to 8\% {\it higher} than the observed value: 
$\alpha_3(m_Z)=0.118\pm 0.003$ \cite{ParticleDataGroup}. 
Now, assuming coupling 
unification, the net (observed) value of $\alpha_3$, 
for the case of MSSM embedded in SU(5) or SO(10), is given by: 
\begin{equation}\label{a4}
  \alpha_3(m_Z)_{\mbox{\scriptsize{net}}}=
	\alpha_3(m_Z)_{\mbox{\scriptsize{MSSM}}}^\circ + 
 \Delta\alpha_3(m_Z)_{\mbox{\scriptsize{DT}}}^{\mbox{\scriptsize{MSSM}}} + 
     \Delta_3'
\end{equation}
where $\Delta\alpha_3(m_Z)_{\mbox{\scriptsize{DT}}}$ and $\Delta_3'$ 
represent GUT-scale threshold corrections respectively due to 
doublet-triplet splitting and the splittings in the other multiplets 
(like the gauge and the Higgs multiplets), all of which are evaluated 
at $m_Z$. Now, owing to mixing between $10_d$ and $16_d$ 
(see Eq. (\ref{a2})), one finds that 
$\Delta\alpha_3(m_Z)_{\mbox{\scriptsize{DT}}}$ is given by 
$(\alpha_3(m_Z)^2/2\pi)(9/7)\ln (M_{\mbox{\scriptsize{eff}}}
\cos\gamma/M_X)$ \cite{BabuWilczekPati}. 

As mentioned above, constraint from proton lifetime sets a lower limit 
on $M_{\mbox{\scriptsize{eff}}}$ given by 
$M_{\mbox{\scriptsize{eff}}}> (1-6)\times 10^{18}$GeV. Thus, even for 
small $\tan\beta\approx 2$ (i.e. $\cos\gamma \approx \tan(\beta/60) 
\approx 1/30$), $\Delta\alpha_3(m_Z)_{\mbox{\scriptsize{DT}}}$ is 
positive; and it increases logarithmically with 
$M_{\mbox{\scriptsize{eff}}}$. Since 
$\alpha_3(m_Z)_{\mbox{\scriptsize{MSSM}}}^\circ$ is higher than 
$\alpha_3(m_Z)_{\mbox{\scriptsize{obs}}}$, and as we saw, 
$\Delta\alpha_3(m_Z)_{\mbox{\scriptsize{DT}}}$ is positive, it follows 
that the corrections due to {\it other} multiplets denoted by 
$\delta_3'=\Delta_3'/\alpha_3(m_Z)$ should be appropriately negative 
so that $\alpha_3(m_Z)_{\mbox{\scriptsize{net}}}$ would agree with 
the observed value. 

In order that coupling unification may be regarded as a natural 
prediction of SUSY unification, as opposed to being a mere coincidence, 
it is important that the magnitude of the net other threshold 
corrections, denoted by $\delta_3'$, be negative but not any more 
than about 8 to 10\% in magnitude (i.e. $-\delta_3' \leqslant (8-10)\%$). 
It was shown in Ref.\cite{BabuWilczekPati} that the contributions from 
the gauge and the minimal set of Higgs multiplets (i.e. $45_H, 16_H, 
\overline{16}_H$ and $10_H$) leads to 
threshold correction, denoted by $\delta_3'$, 
which has in fact a negative sign and quite naturally a magnitude of 
4 to 8\%, 
as needed to account for the observed coupling unification. The 
correction to $\alpha_3(m_Z)$ due to Planck scale physics through the 
effective operator $F_{\mu\nu}F^{\mu\nu}45_H/M$ does not alter the 
estimate of $\delta_3'$ because it vanishes due to antisymmetry in 
the SO(10)- contraction.
 
Imposing that $\delta_3'$ (evaluated at $m_Z$)be negative and 
not any more than about 
10-11\% in magnitude in turn provides a restriction 
on how big the correction due to doublet-triplet splitting - i.e. 
$\Delta\alpha_3(m_Z)_{\mbox{\scriptsize{D$\bar{T}$}}}$ - can be.  That in 
turn sets an upper limit on $M_{eff}\cos\gamma$, and thereby on 
$M_{eff}$ for a given $\tan\beta$. For instance, for MSSM, 
with $\tan\beta=(2,3,8)$, one obtains (see Ref.\cite{BabuWilczekPati}): 
$M_{eff}\le(4,2.66,1)\times 10^{18}$GeV. Thus, conservatively, 
taking $\tan\beta\ge 3$, one obtains: 
\begin{equation}
  M_{eff} \lesssim 2.7\times 10^{18} \mbox{GeV (MSSM)}.
\end{equation} 

\subsection*{Limit on $M_{eff}$ For The case of ESSM }

Next consider the restriction on $M_{eff}$ that would arise for 
the case of the extended supersymmetric standard model (ESSM), which 
introduces an extra pair of vector-like families ($16+\bar{16})$ of 
SO(10)) at the TeV scale \cite{BabuJi}(see also footnote 11). In this 
case, $\alpha_{\mbox{\scriptsize{unif}}}$ is raised to 
0.25 to 0.3, compared to 0.04 
in MSSM. Owing to increased two-loop effects the scale of unification 
M$_X$ is raised to $(1/2-2)\times 10^{17}$GeV, while 
$\alpha_3(m_Z)_{\mbox{\scriptsize{ESSM}}}^\circ$ is lowered to about 
0.112-0.118 \cite{BabuJi,KoldaRussell}. 

With raised M$_X$, the product $M_{eff}\cos\gamma \approx 
M_{eff}(\tan\beta)/60$ can be higher by almost a factor of five 
compared to that for MSSM, without altering 
$\Delta\alpha_3(m_Z)_{\mbox{\scriptsize{DT}}}$. 
Furthermore, since $\alpha_3(m_Z)_{\mbox{\scriptsize{MSSM}}}^\circ$ 
is typically lower than the observed value of $\alpha_3(m_Z)$ 
(contrast this with the case of ESSM), for ESSM, $M_{eff}$ can be higher 
than that for MSSM by as much as a factor of 2 to 3, without requiring 
an enhancement of $\delta_3'$. The net result is that for ESSM embedded in 
SO(10), $\tan\beta$ can span a wide range from 3 to even 30 (say) and 
simultaneously the upper limit on $M_{eff}$ can vary over the 
range (60 to 6)$\times 10^{18} GeV$, satisfying   
\begin{equation}\label{MeffEq}
    M_{eff}\lesssim (6\times 10^{18} 
    \mbox{GeV})(30/\tan\beta) \mbox{\,\,(ESSM)},
\end{equation}
with the unification-scale threshold 
corrections from ``other'' sources denoted by 
$\delta_3'=\Delta_3'/\alpha_3(m_Z)$ being negative, but no more than 
about 5\% in magnitude. As noted above, such values of $\delta_3'$ 
emerge quite naturally for the minimal Higgs system. Thus, one important 
consequence of ESSM is that by allowing for larger values of 
$M_{eff}$ (compared to MSSM), without entailing larger values 
of $\delta_3'$, it can be perfectly compatible with the limit on  
proton lifetime for almost {\it central values} of the parameters 
pertaining to the SUSY spectrum and the relevant matrix elements 
(see Eq.(40)). Further, larger values of $\tan\beta$ (10 to 30, say) 
can be compatible with proton lifetime only for the case of ESSM, but not 
for MSSM. These features are discussed in the text, and also 
exhibited in Table 2.  

\newpage
\begin{center}
  {\bf TABLE 2. VALUES OF PROTON LIFETIME 
	$\left(\Gamma^{-1}(p\rightarrow\bar{\nu}K^+)\right)$ 
	FOR A WIDE RANGE OF PARAMETERS }
\end{center}

\noindent
\begin{tabular}{|c|c|c||c|c||c|}
\hline
\rule[-3mm]{0mm}{8mm}
Parameters  & \multicolumn{2}{|c||}{MSSM $\to$ SO(10)} 
	    & \multicolumn{2}{|c||}{ESSM $\to$ SO(10)}
	    & 							\\
(spectrum/Matrix  &  \multicolumn{2}{|c||}{{\bf Std. d=5}}
            &  \multicolumn{2}{|c||}{{\bf Std. d=5}}
	    &  \raisebox{1.5ex}[0pt] {$\left\{\begin{tabular}{c}
	\vspace*{-1ex} MSSM \\ or \vspace*{-0.6ex} \\ ESSM \end{tabular}
		\right\}\to$ G(224)/SO(10)} \\
element)	 & \multicolumn{2}{|c||}
			{Intermed. $\epsilon'$ \& phase$^\dagger$} 
		 & \multicolumn{2}{|c||}
			{Intermed. $\epsilon'$ \& phase$^\dagger$}
		 & {\bf New d=5}$^{\dagger\dagger}$		\\ \cline{2-6} 
	         & $\tan\beta$=$3$ & $\tan\beta$=$20$ 
		 & $\tan\beta$=$5$ & $\tan\beta$=$20$
	         & Independent of $\tan\beta$ 		\\ \hline 	
Nearly ``central'' & $0.7\times10^{32}$  & 1.6$\times 10^{30}$  
		 & 1.1$\times 10^{34}$  & 0.7$\times10^{33}$  
		 & $0.7\times 10^{33}$ 				\\
\{$\,\,\,$ \}=2	 & yrs & yrs & yrs$^{\bbox{*}}$ 
		 & yrs & yrs$^{\dagger\dagger}$	\\ \hline
Intermediate     & 2.8$\times 10^{32}$  & 0.6$\times 10^{31}$  
		 & 0.4$\times 10^{35}$  & 2.8$\times 10^{33}$  
		 & 2.8$\times 10^{33}$ 			\\
\{$\,\,\,$ \}=8	 & yrs & yrs & yrs$^{\bbox{*}}$ 
		 & yrs & yrs$^{\dagger\dagger}$	\\ \hline
Nearly Extreme   & 1.1$\times 10^{33}$  & 2.6$\times 10^{31}$  
		 & 1.7$\times 10^{35}$  & 1.1$\times 10^{34}$  
		 & 1.1$\times 10^{34}$ 				\\
\{$\,\,\,$ \}=32 & yrs & yrs & yrs$^{\bbox{*}}$ 
		 & yrs & yrs$^{\dagger\dagger}$	\\ \hline
\end{tabular}
$\bbox{^*}${\bf In this case, lifetime is given by the last column.} 

\vspace*{12pt}
$\bullet$ Since we are interested in exhibiting expected proton lifetime 
near the upper end, we are not showing entries corresponding to 
values of the parameters for the SUSY spectrum and the matrix 
element (see Eq.(\ref{e39}), for which the curly bracket appearing 
in Eqs.(\ref{e46}), (\ref{e49-2}), (\ref{e52})) would be less than 
one (see however Table 1). 
In this context, we have chosen here ``nearly central'', 
``intermediate'' and ``nearly extreme'' values of the 
parameters  such that the said curly bracket is given by 2, 8 and 32 
respectively, instead of its extreme upper-end value of 64. For instance, 
the curly bracket would be 2 if $\beta_H=(0.0117)$ GeV$^3$, 
$m_{\tilde{q}}\approx 1.2$ TeV and 
$m_{\tilde{W}}/m_{\tilde{q}}\approx (1/7.2)$, while it would be 8 
if $\beta_H=0.010$ GeV$^3$, 
$m_{\tilde{q}}\approx 1.44$ TeV and 
$m_{\tilde{W}}/m_{\tilde{q}}\approx 1/10$; and it would be 32 if, 
for example, $\beta_H=0.007$ GeV$^3$, 
$m_{\tilde{q}}\approx \sqrt{2}(1.2$ TeV) and 
$m_{\tilde{W}}/m_{\tilde{q}}\approx 1/12$. 

$\bbox{\dagger}$ All the entries for the standard d=5 operators correspond 
to taking an intermediate value of $\epsilon'\approx (1$ to 
$1.4)\times 10^{-4}$ (as opposed to the extreme values of 
$2\times 10^{-4}$ and zero for 
cases I and II, see Eq.(\ref{e33})) and an intermediate phase-dependent 
factor such that the uncertainty factor in the square bracket appearing 
in Eqs.(\ref{e46}) and (\ref{e49-2}) is given by 5, instead of its 
extreme values of $2\times 4=8$ and $2.5\times 4=10$, respectively. 

$\bbox{\dagger\dagger}$ For the new operators, 
the factor [8-1/64] appearing in 
Eq.(\ref{e52}) is taken to be 6, and $K^{-2}$, defined in 
Sec.\ref{Expectations}A, is taken to be 9, which are 
quite plausible, in so far as we wish to obtain reasonable values 
for proton lifetime at the upper end. 

$\bullet$ The standard d=5 operators for both MSSM and ESSM are 
evaluated by taking the upper limit on $M_{eff}$ (defined in the 
text) that is allowed by the requirement of natural coupling 
unification. This requirement restricts threshold corrections and 
thereby sets an upper limit on $M_{eff}$, for a given $\tan\beta$ 
(see Sec.\ref{Expectations} and Appendix). 

${\bbox \ast}$ For all cases, the standard and the new d=5 operators must 
be combined to obtain the net amplitude. For the three cases of ESSM 
marked with an asterisk, and other similar cases which arise for 
low $\tan\beta\approx 3$ to 6 (say), the 
standard d=5 operators by themselves would lead to proton lifetimes 
typically exceeding $(0.1-0.7)\times10^{35}$ years. 
For these cases, however, the contribution 
from the new d=5 operators would dominate, which quite naturally 
lead to lifetimes in the range of $(10^{33}-10^{34})$ years (see last 
column).

$\bullet$ As shown above, the case of MSSM embedded in SO(10) is 
tightly constrained by present empirical lower limit on proton 
lifetime (Eq.(\ref{e41})). In this case, only low values of 
$\tan\beta\leq 3$, with the parameters (pertaining to the SUSY 
spectrum, matrix element and phase-dependent factor) having their 
``nearly extreme'' or extreme values (as in Eq.(\ref{e39})) can 
lead to lifetimes in the range of $(1-3)\times 10^{33}$yrs (see Table 
and Eq.(\ref{e47})), compatible with present empirical limit. 
Other cases of MSSM - especially with $\tan\beta \geq 5$ and/or 
``nearly central'' or even ``intermediate'' range of parameters - 
seem to be excluded, subject 
(of course) to our requirement for natural coupling unification 
(see Sec.\ref{Expectations} and Appendix).

$\bullet$ Including contributions from the standard and the new 
operators, the case of ESSM, embedded in either G(224) or SO(10), 
is, however, fully consistent with present limits on proton lifetime 
for a wide range of parameters; 
at the same time it provides optimism that proton decay will be 
discovered in the near future, with a lifetime $\leq 10^{34}$ years. 

$\bullet$ The lower limits on proton lifetime are not exhibited. 
In the presence of the new operators, these can typically be as low 
as about $10^{29}$ years (even for the case of ESSM embedded in 
SO(10)). Such limits and even higher are of course long excluded 
by experiments. 

$\bullet$ Allowing for a wide variation in the relevant parameters, 
we thus see that a {\it conservative upper limit} on proton 
lifetime is given by the range of $(1/2-1)\times 10^{34}$ years 
for ESSM and (of course) MSSM, embedded in SO(10) or string-G(224).

\end{document}